\documentclass[11pt,a4paper]{article}
\usepackage{jcappub}

\usepackage{graphicx}
\usepackage{xcolor}
\usepackage{mathrsfs,mathtools}
\usepackage{physics,amssymb}
\usepackage{bm}
\usepackage{soul}
\usepackage{cancel}
\usepackage{xurl}
\usepackage{longtable}
\usepackage{xspace}
\usepackage{acronym}
\usepackage{booktabs}
\usepackage{tabularx}
\usepackage{subcaption}
\newcolumntype{C}{>{\centering\arraybackslash}X}

\hypersetup{colorlinks=true
,urlcolor=DARKBLUE
,anchorcolor=DARKBLUE
,citecolor=DARKBLUE
,filecolor=DARKBLUE
,linkcolor=DARKBLUE
,menucolor=DARKBLUE
,linktocpage=true
,pdfproducer=medialab
,pdfa=true
}

\newcommand{\ee}{\mathrm{e}}
\newcommand{\Mpl}{M_\mathrm{Pl}}
\newcommand{\As}{A_{\mathrm{s}}}
\newcommand{\ns}{n_{\mathrm{s}}}

\newcommand{\gO}{\mathrm{O}}
\newcommand{\gZ}{\mathrm{Z}}
\newcommand{\water}{\text{water}}
\newcommand{\peak}{\text{peak}}
\newcommand{\uth}{\mathrm{th}}

\newcommand{\uc}{\mathrm{c}}
\newcommand{\uD}{\mathrm{D}}

\newcommand{\uf}{\mathrm{f}}

\newcommand{\sfH}{\mathsf{H}}

\newcommand{\ui}{\mathrm{i}}

\newcommand{\bfk}{\mathbf{k}}

\newcommand{\calN}{\mathcal{N}}
\newcommand{\scrN}{\mathscr{N}}
\newcommand{\bfn}{\mathbf{n}}

\newcommand{\calP}{\mathcal{P}}

\newcommand{\ur}{\mathrm{r}}

\newcommand{\bfx}{\mathbf{x}}

\newcommand{\beae}[1]{\begin{equation}\begin{aligned} #1 \end{aligned}\end{equation}}

\newcommand{\bae}[1]{\begin{align} #1 \end{align}}
\newcommand{\bce}[1]{\begin{cases} #1 \end{cases}}
\newcommand{\bde}[1]{\begin{dcases} #1 \end{dcases}}

\newcommand{\relmiddle}[1]{\mathrel{}\middle#1\mathrel{}}

\definecolor{MONZA}{HTML}{CF000F}
\definecolor{DARKBLUE}{HTML}{00008b}
\definecolor{DARKMAGENTA}{HTML}{8b008b}

\acrodef{CMB}{cosmic microwave background}
\acrodef{PBH}{primordial black hole}
\newacroplural{PBH}{primordial black hole}
\acrodef{MD1}{mass-dimension $1$}
\acrodef{VEV}{vacuum expectation value}
\acrodef{PDF}{probability density function}
\acrodef{STOLAS}{STOchastic LAttice Simulation}
\acrodef{EoM}{equation of motion}
\newacroplural{EoM}{equations of motion}

\begin{document}
\title{STOchastic LAttice Simulation of hybrid inflation}
\date{\today}

\collaborationImg{\includegraphics[width=0.3\hsize]{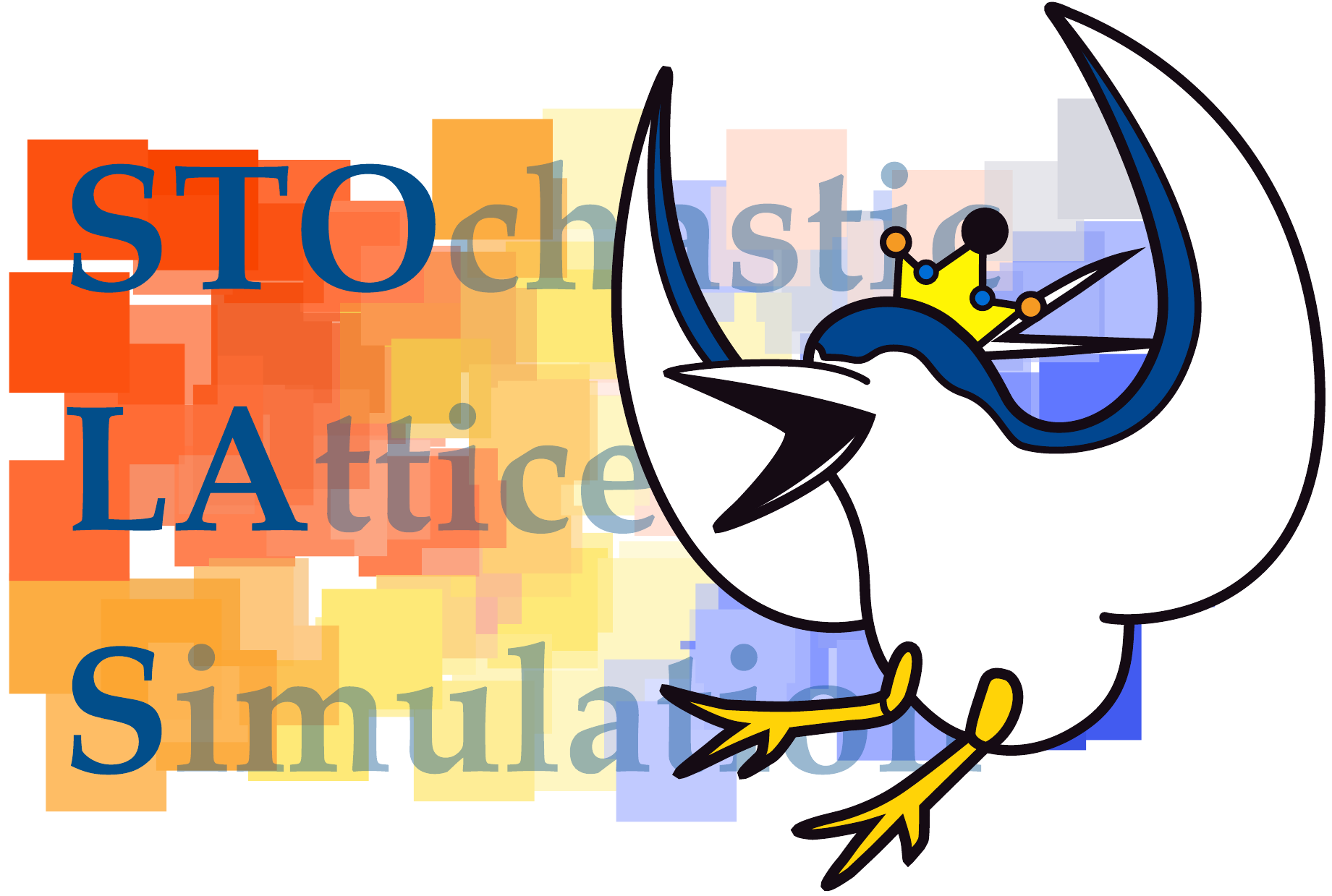}}
\author[a,b]{Tomoaki Murata}
\author[b]{and Yuichiro Tada}

\affiliation[a]{Department of General Education, Tokyo Metropolitan College of Industrial Technology, 
Shinagawa, Tokyo 140-0011, Japan}
\affiliation[b]{Department of Physics, Rikkyo University, Toshima, Tokyo 171-8501, Japan}

\emailAdd{tomoaki-m@metro-cit.ac.jp}
\emailAdd{yuichiro.tada@rikkyo.ac.jp}

\abstract{
We investigate the spatial profile of the curvature perturbation generated in multi-waterfall hybrid inflation models, which are known to produce various topological defects. 
Using the lattice simulation code \acl{STOLAS}, based on the stochastic formalism of inflation, we analyse six cases by varying the number of waterfall fields $n$ and the functional form of the inflaton potential (``Quadratic'' and ``Cubic'' cases). 
Our statistical analysis shows that the \acp{PDF} and power spectra are broadly consistent with the so-called stochastic-$\delta\calN$ algorithm. 
The ``Cubic'' case also exhibits a characteristic upper bound in the \ac{PDF}, as discovered in our previous work, that suppresses \acl{PBH} formation while potentially affecting halo formation. 
Furthermore, we employ the Euler characteristic as a topological diagnostic tool to identify the structures of the waterfall fields as well as the curvature perturbation.
We find that the topological defects, such as domain walls ($n=1$), cosmic strings ($n=2$), and monopoles ($n=3$), are reconnected during inflation into finer structures by the stochastic noise, making their correlation lengths much smaller than the Hubble scale at the critical point of the waterfall phase transition counterintuitively. 
The Euler characteristic also implies global structures of the curvature perturbation for $n=1$, though we do not conclude if they are due to the domain wall, because neither the strings ($n=2$) nor monopoles ($n=3$) leave such structures. 
The global structures of the curvature perturbation will provide a novel probe for the physics of the early universe.
}
\arxivnumber{}

\noindent RUP-26-3

\maketitle
\flushbottom

\section{Introduction}

The inflationary paradigm provides a well-established mechanism for solving the horizon and flatness problems and for generating the primordial perturbations that later grew into the observed large-scale structure~\cite{Starobinsky:1980te,Sato:1981qmu,Guth:1980zm,Linde:1981mu,Albrecht:1982wi,Linde:1983gd}.
While much attention has traditionally focused on the nearly scale-invariant, small-amplitude perturbations observed in the \ac{CMB}~\cite{Planck:2018jri}, there is growing interest in inflationary scenarios that produce large curvature perturbations on small scales, since such perturbations can potentially collapse to form \acp{PBH}~\cite{Hawking:1971ei,Carr:1974nx,Carr:1975qj} (see also Refs.~\cite{Carr:2020gox,Escriva:2022duf,Yoo:2022mzl,Carr:2023tpt} for several reviews), offer a complementary probe of early Universe physics, and motivate detailed studies of mechanisms that can amplify curvature perturbations.

Though a single real scalar field is often supposed to drive inflation in a minimal setup, multiple fields can be relevant in general.
Hybrid inflation~\cite{Linde:1993cn} serves as a representative example in which multiple scalar fields play essential roles in both driving and terminating inflation.
In this model, the so-called \emph{inflaton} field governs the slow-roll expansion along a nearly flat potential valley until it reaches a critical point where the effective mass of the auxiliary \emph{waterfall} fields becomes negative.
This tachyonic instability marks the onset of the waterfall phase, during which the waterfall field rapidly rolls down its potential and inflation comes to an end. 
Since the dynamics of the waterfall phase depend sensitively on the field fluctuations at the critical point, the system is intrinsically non-perturbative, often leading to a significant enhancement of curvature perturbations.
Such enhancements can naturally produce \acp{PBH}, particularly in the mild-waterfall regime, where the waterfall stage lasts for several e-folds~\cite{Garcia-Bellido:1996mdl,Clesse:2010iz,Lyth:2010zq,Kodama:2011vs,Mulryne:2011ni,Bugaev:2011qt,Bugaev:2011wy,Lyth:2012yp,Clesse:2012dw,Guth:2012we,Clesse:2015wea}.

For such regimes, stochastic formalism provides a useful complementary description.
In this approach, the dynamics of long-wavelength (IR) modes are effectively described by classical stochastic equations, which are derived by integrating out the short-wavelength (UV) modes that continually cross the Hubble horizon~\cite{Starobinsky:1982ee,Starobinsky:1986fx,Nambu:1987ef,Nambu:1988je,Kandrup:1988sc,Nakao:1988yi,Nambu:1989uf,Mollerach:1990zf,Linde:1993xx,Starobinsky:1994bd,Finelli:2008zg}.
The UV modes act as stochastic noise sourcing the IR fields, leading to a Langevin-type equation for the coarse-grained inflaton fields. 
This formalism captures the effect of quantum fluctuations during inflation and provides a non-perturbative framework to study the evolution of curvature perturbations on superHubble scales.
In conjunction with the $\delta N$ approach~\cite{Starobinsky:1985ibc,Salopek:1990jq,Sasaki:1995aw,Sasaki:1998ug,Wands:2000dp,Lyth:2004gb,Lyth:2005fi}, the formalism can calculate the conserved curvature perturbation $\zeta$ non-perturbatively as a fluctuation $\delta\calN$ in the stochastic first passage e-folding number $\calN$ to a certain end surface, known as the stochastic-$\delta\calN$ formalism~\cite{Fujita:2013cna,Fujita:2014tja,Vennin:2015hra}.
It enables us to quantitatively estimate the power spectrum of the curvature perturbation in mild-waterfall hybrid inflation~\cite{Kawasaki:2015ppx,Tada:2023pue}.

As a further development, we proposed a real-space lattice simulation of inflation based on the stochastic formalism with Mizuguchi, dubbed \emph{\ac{STOLAS}}~\cite{Mizuguchi:2024kbl}.
\ac{STOLAS} is the C++ code that can directly produce three-dimensional maps of the curvature perturbation. 
This approach allows us to examine, in a single consistent setup, both the probability distribution of large curvature fluctuations and their spatial correlations.
One main purpose of this work is to clarify the previous stochastic-$\delta\calN$ formalism by comparing its power spectrum with a simple Fourier transform of the lattice result.

The topological configuration of the curvature perturbation is also an interesting subject, as hybrid inflation is known to form topological defects, such as the domain wall, cosmic string, and monopole, in general.
While these defects are ultimately formed near the end of inflation, their seeds can be generated gradually during the waterfall transition as field fluctuations grow.
Such inhomogeneous field configurations may influence the structure of the curvature perturbation.
\ac{STOLAS} with the $\delta N$ formalism, a non-perturbative lattice simulation of the curvature perturbation, can capture this effect, as the other main topic of this paper.

The paper is organised as follows.
In Sec.~\ref{sec: review STOLAS}, we review the scheme of \ac{STOLAS}.
Section~\ref{sec: review hybrid} is devoted to a review of the hybrid inflation scenario and its key features.
In Sec.~\ref{sec: PDF and PS}, we present the statistical properties of the curvature perturbation obtained within \ac{STOLAS}.
In Sec.~\ref{sec: defect}, we investigate the topological feature of the field configuration of the waterfall fields and the curvature perturbation.
Section~\ref{sec: conclusion} is devoted to conclusions.

\section{Review of STOchastic LAttice Simulation} \label{sec: review STOLAS}

In this section, we briefly review \ac{STOLAS}~\cite{Mizuguchi:2024kbl}. Several extensions are also implemented to address hybrid inflation.
Our code is available at \url{https://github.com/STOchasticLAtticeSimulation/STOLAS_dist/tree/main/hybrid-inflation}.

\subsection{Discrete stochastic formalism of inflation}

While the original \ac{STOLAS} code in Ref.~\cite{Mizuguchi:2024kbl} is implemented for single-field models, we extend it to the canonical multi-field case, where the \ac{EoM} and the Friedmann constraint are given by
\bae{\label{eq: single noise EoM}
    \bde{
        \dv{\phi_i(N,\bfx)}{N} = \frac{\pi_i(N,\bfx)}{H(N,\bfx)} + \calP^{1/2}_{ij}(N,\bfx)\xi_{j}(N,\bfx), \\
        \dv{\pi_i(N,\bfx)}{N} = - 3 \pi_i(N,\bfx) - \frac{V_i\qty(\bm{\phi}(N,\bfx))}{H(N,\bfx)}, \\
        3\Mpl^2H^2(N,\bfx)=\frac{\pi_i^2(N,\bfx)}{2}+V\qty(\bm{\phi}(N,\bfx)).
    }
}
Here, $\bm{\phi}=(\phi_i)_{i=0,\dots,n}$ are scalar fields and $\bm{\pi}=(\pi_i)_{i=0,\dots,n}$ are their conjugate momenta.\footnote{Strictly speaking, $\pi_i$ is not the conjugate momenta with respect to $N$, but it is rather related to the cosmic time derivative. Even the first equation of \eqref{eq: single noise EoM} can be understood as its definition. This definition simplifies the last Friedmann equation. See, e.g., Ref.~\cite{Pinol:2020cdp} for more details.}
The e-folding number $N$ is used as the time variable, and $V_i=\partial_{\phi_i}V$ is the derivative of the scalar potential $V(\bm{\phi})$. 
In the stochastic formalism, all fields are understood as the superHubble coarse-grained parts. Accordingly, they receive random fluctuations due to the continuous horizon crossing of subHubble modes, represented by the $\xi_i$ term.
The normalised white noise $\xi_i$ satisfies
\bae{
    \expval{\xi_i(N,\bfx)}=0 \qc \expval{\xi_i(N,\bfx)\xi_j(N',\bfx')}=\delta_{ij}\delta(N-N')\frac{\sin k_\sigma(N)\abs{\bfx-\bfx'}}{k_\sigma(N)\abs{\bfx-\bfx'}},
}
where $k_\sigma(N)$ represents the comoving coarse-graining scale of the stochastic formalism.
In this paper, the angle brackets $\expval{\cdot}$ denote the stochastic average.
The physical coarse-graining scale $k_\sigma(N)/a(N)$, where $a(N)\propto\ee^N$ is the scale factor, is a constant as a model parameter (or a simulation setup in \ac{STOLAS}), 
and should be sufficiently smaller than the Hubble parameter $H(N,\bfx)$ in any typical realisation to ensure that all fields are well superHubble.
In this work, we adopt
\bae{
    k_\sigma(N)=\sigma a(N)\sfH \qc \sigma=\frac{1}{16}, 
}
where the scale factor $a(N)$ and the characteristic scale $\sfH$ are normalised at the simulation onset $N=0$ by
\bae{
    a(N=0)\sfH L=2\pi,
}
with the (comoving) simulation box size $L$. In the lattice simulation, only the dimensionless wavenumber
\bae{
    n_\sigma(N)=\frac{k_\sigma(N)L}{2\pi}=\sigma\ee^N,
}
appears, and hence the box size $L$ can be arbitrarily fixed after the simulation. One should then monitor whether the coarse-graining scale $k_\sigma$ with the defined $L$ is well superHubble at any grid throughout the simulation.
In Eq.~\eqref{eq: single noise EoM}, we neglected the noise for $\pi_i$ as it is subdominant in the slow-roll regime.
We also adopt the slow-roll approximation for the noise amplitude $\calP_{ij}^{1/2}$ as
\bae{
    \calP_{ij}^{1/2}(N,\bfx)\simeq\frac{H(N,\bfx)}{2\pi}\delta_{ij}.
}

These stochastic equations can be implemented by a numerical lattice simulation with a certain appropriate discretisation.
In the temporal direction, the \ac{EoM} is discretised in the Euler--Maruyama way as
\bae{\label{eq: discrete EoM}
    \bde{
        \phi_i(N+\Delta N,\bfx)-\phi_i(N,\bfx)=\Delta_\uD\phi_i(N,\bfx)+\frac{H(N,\bfx)}{2\pi}\Delta W_i(N,\bfx), \\
        \pi_i(N+\Delta N,\bfx)-\pi_i(N,\bfx)=\Delta_\uD\pi_i(N,\bfx),
    }
}
where $\Delta_\uD\phi_i$ and $\Delta_\uD\pi_i$ stand for the deterministic parts of the equations,
\bae{
    \Delta_\uD\phi_i(N,\bfx)=\frac{\pi_i(N,\bfx)}{H(N,\bfx)}\Delta N \qc 
    \Delta_\uD\pi_i(N,\bfx)=-\pqty{3\pi_i(N,\bfx)+\frac{V_i\qty(\bm{\phi}(N,\bfx))}{H(N,\bfx)}}\Delta N,
}
with the time step $\Delta N$, while $\Delta W_i$ is a spatially-correlated Gaussian random variable satisfying
\bae{\label{eq: DW correlation}
    \expval{\Delta W_i(N,\bfx)}=0 \qc 
    \expval{\Delta W_i(N,\bfx)\Delta W_j(N',\bfx')}=\delta_{ij}\delta_{NN'}\frac{\sin k_\sigma(N)\abs{\bfx-\bfx'}}{k_\sigma(N)\abs{\bfx-\bfx'}}\Delta N.
}
For stability, we adopt the fourth-order Runge--Kutta method to evaluate the deterministic parts $\Delta_\uD\phi_i$ and $\Delta_\uD\pi_i$.

The stochastic equations~\eqref{eq: single noise EoM} do not include any spatial derivative, and hence each spatial point $\bfx$ is independent except for the noise correlation~\eqref{eq: DW correlation}.%
\footnote{Strictly speaking, dropping the spatial derivatives in Eq.~\eqref{eq: single noise EoM} corresponds to taking the leading order in the $\sigma$ expansion, while finite $\sigma$ is retained in the noise amplitude and correlator. This treatment is justified if the results are dominated by the $\sigma^0$ contribution, which is numerically supported by the insensitivity of our results to the choice of $\sigma$ demonstrated in Appendix~\ref{app:sigma}.}
That is, the spatial discretisation is non-trivial only for the generation of $\Delta W_i$.
The real-space correlation~\eqref{eq: DW correlation} implies that its Fourier mode only has the wavenumber $k_\sigma(N)$.
In fact, this spatial correlation can be well approximated by the discrete (inverse) Fourier transform\footnote{Note that our definition of the Fourier transform differs from that in the original paper~\cite{Mizuguchi:2024kbl} by $N_L^3$. Also, our wavenumber $\bfn$ corresponds to their shifted one $\tilde{\bfn}$.}
\bae{
    \Delta W_i(N,\bfx)=\sum_{\bfn}\widetilde{\Delta W_i}(N,\bfn)\ee^{i\frac{2\pi}{L}\bfn\cdot\bfx} \qc \bfn=\left\{(n_x,n_y,n_z)\in\mathbb{Z}^3\relmiddle{|} -\frac{N_L}{2}+1\leq n_i\leq\frac{N_L}{2}\right\},
}
with independent Gaussian random variables $\widetilde{\Delta W}_i(N,\bfn)$ non-vanishing only for $n=\abs{\bfn}\sim n_\sigma(N)$.
Here, $N_L$ is the number of grids in each direction, i.e., the total number of grids is $N_L^3$.
In this paper, we allow $\pm1/2$ tolerance: $\abs{n-n_\sigma(N)}\leq1/2$.
The lattice simulation will be performed while $n_\sigma(N)<N_L/2-1$.
See the original paper~\cite{Mizuguchi:2024kbl} for more details to ensure the reality of $\Delta W_i(N,\bfx)$, etc.

\subsection[Average $\delta\calN$]{\boldmath Average $\delta \calN$} \label{sec: average deltaN}

According to the $\delta N$ formalism~\cite{Starobinsky:1985ibc,Salopek:1990jq,Sasaki:1995aw,Sasaki:1998ug,Wands:2000dp,Lyth:2004gb,Lyth:2005fi}, the conserved curvature perturbation $\zeta$ is equivalent to the fluctuation in the e-folding number from an initial flat hypersurface to a final uniform-density hypersurface.
In the stochastic formalism, it can be calculated as the stochastic fluctuation in the first passage time $\calN$ from the initial field values to the final uniform-density condition, known as the stochastic-$\delta\calN$ approach (see, e.g., Refs.~\cite{Fujita:2013cna,Fujita:2014tja,Vennin:2015hra}).
In other words, the curvature perturbation at the spatial point $\bfx$ is given by
\bae{\label{eq: zeta as dN}
    \zeta(\bfx)=\delta\calN(\bfx)=\calN(\bfx)-\overline{\calN(\bfx)},
}
where $\calN(\bfx)$ is the first passage time realised at $\bfx$, and the simulation box average $\overline{\calN(\bfx)}$ is subtracted so that $\overline{\zeta(\bfx)}=0$.

The final hypersurface is well approximated by the end of inflation, as often done in the literature.
In the original paper~\cite{Mizuguchi:2024kbl}, the first passage time $\calN(\bfx)$ is calculated by numerically solving the \ac{EoM}~\eqref{eq: discrete EoM} from the field values at the end of the lattice simulation to the end of inflation without the noise term to focus on the fluctuations during the lattice simulation, supposing that the noise effects after the lattice simulation are negligible.
However, in the mild-waterfall hybrid inflation, as our interest, stochastic effects remain significant until near the end of inflation.
We therefore continue to include the noise term even after the lattice simulation, but neglect its spatial correlations, since the lattice sites are separated beyond the coarse-graining scale.
Specifically, the independent noise is included until $\eta_\ur=-1.2$ while the final hypersurface is defined by $\eta_\ur=-2$, where $\eta_\ur$ is the slow-roll parameter defined by Eq.~\eqref{eq: etar} below.

The first passage time calculated in this way is related to the curvature perturbation coarse-grained over $k_\sigma^{-1}$ at $\eta_\ur=-1.2$, which is much smaller than the lattice grid size.
Such small-scale fluctuations unresolved by the lattice cause numerical artefacts in the discrete Fourier transform.
We would rather obtain the curvature perturbation coarse-grained over the lattice scale.
To this end, at each grid point, we repeat the above uncorrelated-noise stage multiple times with different realisations of the noise, and estimate the mean value of $\calN(\bfx)$ over these realisations.
Namely, the coarse-grained curvature perturbation $\zeta_\uc(\bfx)$ is calculated as
\bae{\label{eq: zetac}
    \zeta_\uc(\bfx)=\expval{\calN\pqty{\bm{\phi}(N_\uf,\bfx),\bm{\pi}(N_\uf,\bfx)}}-\overline{\expval{\calN\pqty{\bm{\phi}(N_\uf,\bfx),\bm{\pi}(N_\uf,\bfx)}}},
}
where $\bm{\phi}(N_\uf,\bfx)$ and $\bm{\pi}(N_\uf,\bfx)$ are the field and momentum values at the end of the lattice simulation $N_\uf$ at the spatial point $\bfx$, and $\calN(\bm{\phi})$ is the first passage time from these configuration to the final hypersurface.
In this paper, we evaluate the stochastic average $\expval{\cdot}$ by $20$ realisations of the uncorrelated noise at each grid point.
We hereafter drop the subscript `$\uc$' for conciseness.
See Appendix~\ref{sec: zoom in} for the validity of this averaging procedure.

\section{Hybrid inflation}%
\label{sec: review hybrid}

In this section, we review the aspects of the mild-waterfall hybrid inflation.
Hybrid inflation, originally proposed by Linde~\cite{Linde:1993cn}, is driven by two types of scalar fields: the inflaton and the waterfall fields.
In this paper, we employ a single inflaton $\phi$ ($=\phi_0$ in the previous notation) and multiple waterfall fields $\bm{\psi}=\pqty{\psi_1,\psi_2,\cdots,\psi_n}$ ($=(\phi_i)_{i=1,\dots,n}$) with general $n$.
The potential respects $\gO(n)$ (or $\gZ_2$ for $n=1$) symmetry as given by
\bae{
    V(\phi,\bm{\psi})=\Lambda^4\bqty{\pqty{1-\frac{\psi_\ur^2}{M^2}}^2+2\frac{\phi^2\psi_\ur^2}{\phi_\uc^2M^2}+v(\phi)},
}
where $\psi_\ur = \sqrt{\sum_{i=1}^n\psi_{i}^2}$ is the radial direction of waterfall fields, and $\Lambda$, $M$, and $\phi_\uc$ are dimensionful model parameters.
The unstable origin of the waterfall fields, $\bm{\psi}=\bm{0}$, can be stabilised by the coupling to the inflaton $\phi$ when $\phi$ is larger than the critical point $\phi_\uc$.
Hence, hybrid inflation consists of two phases. 
The inflaton first rolls down its potential $v(\phi)$ along the valley $\bm{\psi}\sim\bm{0}$ during $\phi>\phi_\uc$ (we suppose $\phi>0$ throughout this work without loss of generality).
After reaching the critical point $\phi_\uc$, the waterfall fields are destabilised and roll down toward the true minimum $\psi_\ur=M$, ending inflation.
Depending on the parameter choice, the waterfall direction can be sufficiently flat so that inflation continues for several e-folds even after the critical point.
In particular, the model where the waterfall phase lasts for more than a few e-folds but less than $\sim50$ e-folds corresponding to the current horizon scale, is called the mild-waterfall hybrid inflation~\cite{Garcia-Bellido:1996mdl,Lyth:2010zq,Bugaev:2011qt,Bugaev:2011wy,Lyth:2012yp,Guth:2012we,Halpern:2014mca,Clesse:2015wea,Kawasaki:2015ppx,Tada:2023pue,Tada:2023fvd,Tada:2024ckk,Murata:2025onc}.
In this model, the waterfall dynamics, such as its duration and the direction of the final \ac{VEV}, $\bm{\psi}/\abs{\bm{\psi}}$, are highly sensitive to the field fluctuations around the critical point, indicating its non-perturbative nature.
We hereby series-expand the inflaton potential $v(\phi)$ around the critical point for a general analysis as
\bae{
    v(\phi) = \frac{\phi - \phi_\uc}{\mu_{1}}
    - \frac{(\phi - \phi_\uc)^2}{\mu_{2}^2}+\frac{(\phi-\phi_\uc)^3}{\mu_{3}^3}+\cdots,
    \label{eq:potential}
}
with model parameters $\mu_{1}$, $\mu_{2}$, $\mu_{3}$, and so on (the negative sign of the quadratic term is for a later convenience).
We define the end surface both for inflation and the $\delta N$ formalism by $\eta_\ur=-2$ with
\bae{\label{eq: etar}
    \eta_\ur=\Mpl^2\frac{V_{\psi_\ur\psi_\ur}}{V},
}
for simplicity as done in the literature~\cite{Kodama:2011vs,Clesse:2015wea,Murata:2025onc}.
Throughout this paper, we neglect the stochastic noise for $\phi$, which is almost irrelevant to the waterfall dynamics, compared to the noise for $\bm{\psi}$.

Some of the above parameters are constrained by the \ac{CMB} observations.
Suppose that the corresponding perturbations crossed the horizon during the valley phase.
The waterfall fields are irrelevant to them, and the amplitude $\As$ and the spectral index $\ns$ of the power spectrum of the curvature perturbation are given by
the textbook formulae in the (quasi) single-field, slow-roll models as
\beae{\label{eq: CMB const.}
    &\As=\frac{1}{24\pi^2\Mpl^4}\frac{V}{\epsilon_\phi} \simeq \frac{\Lambda^4\mu_1^2}{12\pi^2\Mpl^6}, \\
    &\ns=1-6\epsilon_\phi+2\eta_\phi \simeq 1-\frac{4\Mpl^2}{\mu_2^2}+\frac{12\Mpl^2(\phi_*-\phi_\uc)}{\mu_3^3}+\cdots,
}
where $\epsilon_\phi$ and $\eta_\phi$ are the slow-roll parameters in the $\phi$'s direction defined by
\bae{
    \epsilon_\phi=\frac{\Mpl^2}{2}\pqty{\frac{v'(\phi)}{v(\phi)}}^2 \qc
    \eta_\phi=\Mpl^2\frac{v''(\phi)}{v(\phi)}.
}
While we only kept the leading term in the first line in Eq.~\eqref{eq: CMB const.}, the subleading term is also shown in the second line, where $\phi_*$ is the field value at the time when the \ac{CMB}-scale perturbations crossed the horizon.
The \ac{CMB} constraints $\As=2.1\times 10^{-9}$ and $\ns=0.965$~\cite{Planck:2018vyg} then fix certain parameter combinations.
In particular, if the cubic or higher terms in $v(\phi)$~\eqref{eq:potential} are negligible as usual hierarchies, all parameters can be fixed a priori without solving the dynamics to determine $\phi_*$. 
We call this case the ``Quadratic" model.
On the other hand, in the ``Cubic" model where the cubic term in $v(\phi)$ is relevant with a mild tuning among the parameters, the parameters are randomly chosen first, the background dynamics are solved to find $\phi_*$, and then only the parameter choices satisfying the constraint~\eqref{eq: CMB const.} are allowed a posteriori.
Throughout the work, we fix the characteristic field values by $M = \phi_{\uc}/\sqrt{2} = 0.01\Mpl$~\cite{Murata:2025onc}.

\subsection{Fitting formula of power spectrum}

The standard linear perturbation theory for the curvature perturbation breaks down around the critical point because of its non-perturbative nature that the field fluctuations determine the background dynamics during the waterfall phase.
The semi-perturbative analysis developed in Refs.~\cite{Kodama:2011vs,Clesse:2013jra,Clesse:2015wea} suggest the following fitting formula of the power spectrum $\calP_\zeta(k)=\frac{k^3}{2\pi^2}\int\dd[3]{\bfx}\expval{\zeta(\bfx)\zeta(\mathbf{0})}\ee^{-i\bfk\cdot\bfx}$ around the critical point~\cite{Tada:2023pue}:
\bae{
    \calP_{\zeta} (k) = \calP_{\zeta}^{(\rm peak)}
\exp[-2 \frac{\chi_{2}}{\Delta N_{1}^2}
\qty((N_{\rm water} - \calN_{k})^2 + \frac{2 \Mpl^2}{3\mu_{2}^2} (N_{\rm water} - \calN_{k})^3)],
\label{eq:analytic_formula}
}
with the fitting parameters $N_\water$ and $\calP_\zeta^{(\peak)}$.
Here, $\calN_{k}$ is the backward e-folds at $k = k_\sigma$ from the end of inflation, $\chi_{2}$ is defined by
\bae{
\chi_{2} = \frac{1}{2} \ln\qty[\frac{12 \sqrt{2\pi^3} \Mpl^6 \Pi}{n\Lambda^{4} \mu_{1}^{2}}],
}
with the characteristic parameter combination
\bae{
    \Pi=\frac{M\sqrt{\phi_\uc\mu_1}}{\Mpl^2},
}
and $\Delta N_{1}$ is determined from the solution of the following equation:
\bae{
    \chi_{2} = \frac{4}{\Pi^2} \qty(\Delta N_{1}^2 + \frac{2\Mpl^2}{3\mu_{2}^2} \Delta N_{1}^3).
}
This formula has also been confirmed in the stochastic-$\delta\calN$ formalism, a non-perturbative approach to the superHubble fluctuations~\cite{Tada:2023fvd}.
There, instead of using the lattice simulation, the spatially one-point statistics of $\delta\calN$ are derived by repeatedly solving the stochastic equation~\eqref{eq: single noise EoM} without the spatial correlation, and then the power spectrum (i.e., the two-point function of the curvature perturbation) is extracted via the scale-time correspondence $k=k_\sigma(N)$.
This algorithm, proposed by Refs.~\cite{Fujita:2013cna,Fujita:2014tja}, is sometimes referred to as \emph{the stochastic-$\delta\calN$ formalism in the narrow sense}.
One purpose of this paper is to clarify this algorithm by \ac{STOLAS}.

\section{Probability density function and power spectrum}%
\label{sec: PDF and PS}

\begin{figure}
    \centering
    \begin{tabular}{ccc}
        \begin{minipage}{0.3\hsize}
            \centering
            \includegraphics[width=0.95\hsize]{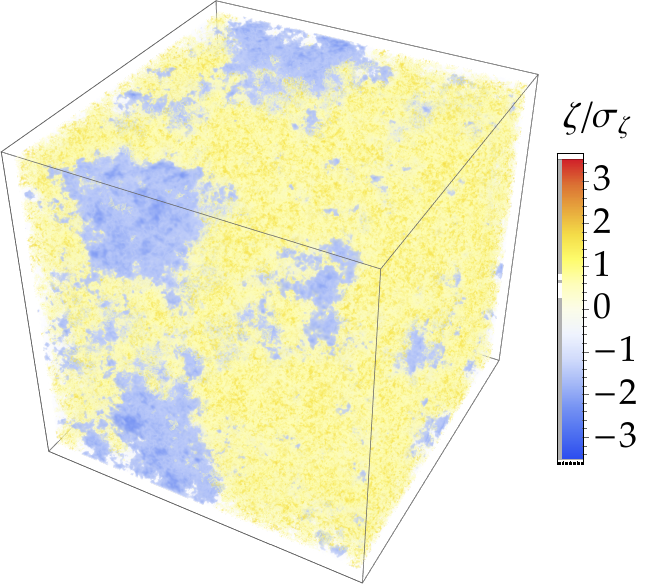}
            \includegraphics[width=0.95\hsize]{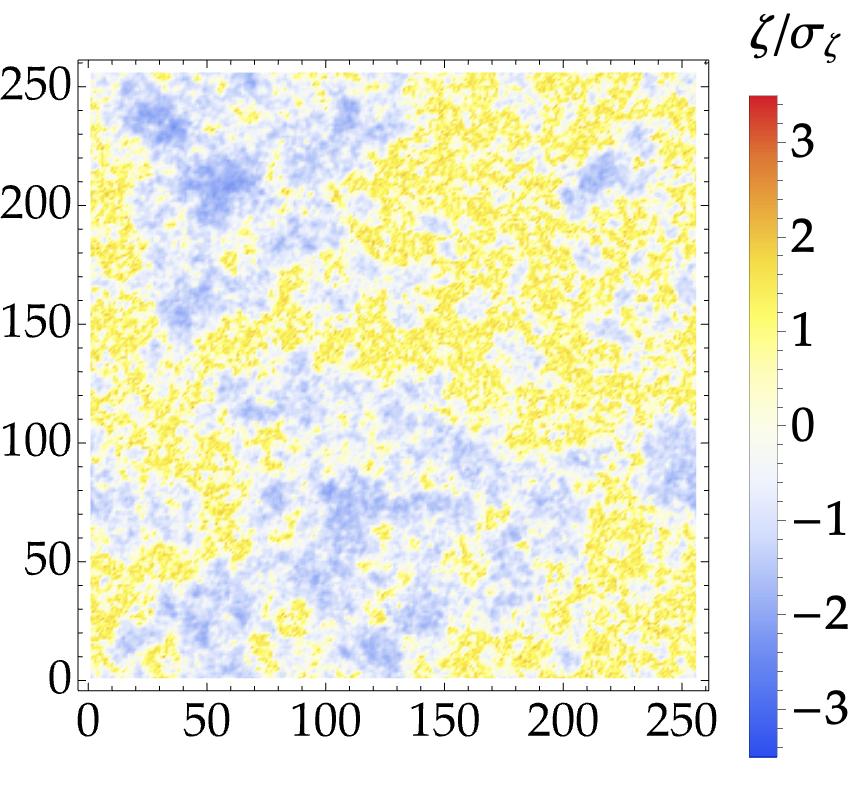}
            \subcaption*{``Quadratic" $n=1$}
        \end{minipage} &
        \begin{minipage}{0.3\hsize}
            \centering
            \includegraphics[width=0.95\hsize]{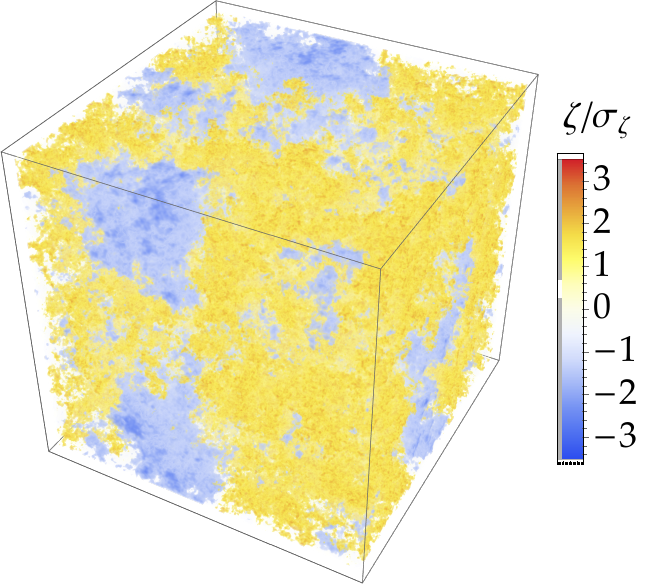}
            \includegraphics[width=0.95\hsize]{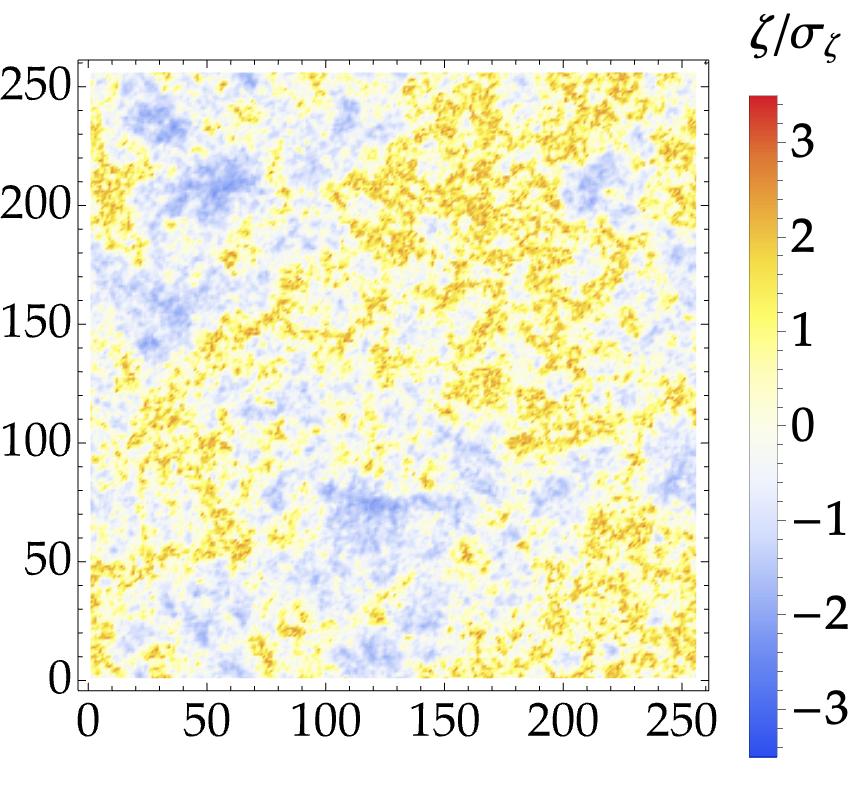}
            \subcaption*{``Quadratic" $n=2$}
        \end{minipage} &
        \begin{minipage}{0.3\hsize}
            \centering
            \includegraphics[width=0.95\hsize]{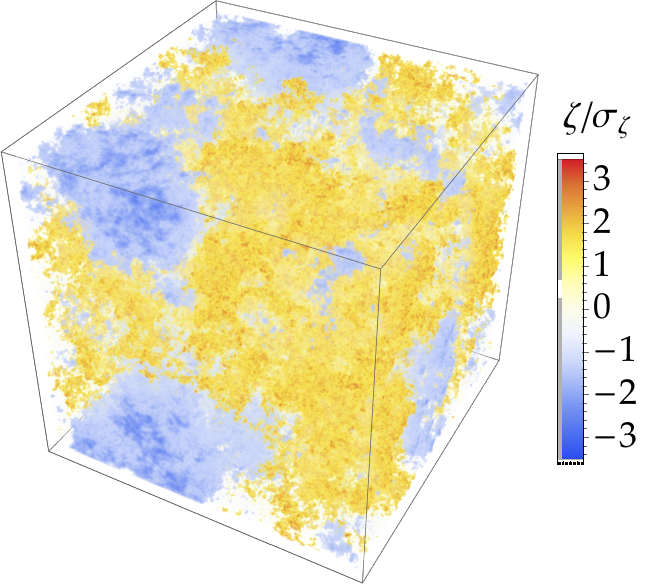}
            \includegraphics[width=0.95\hsize]{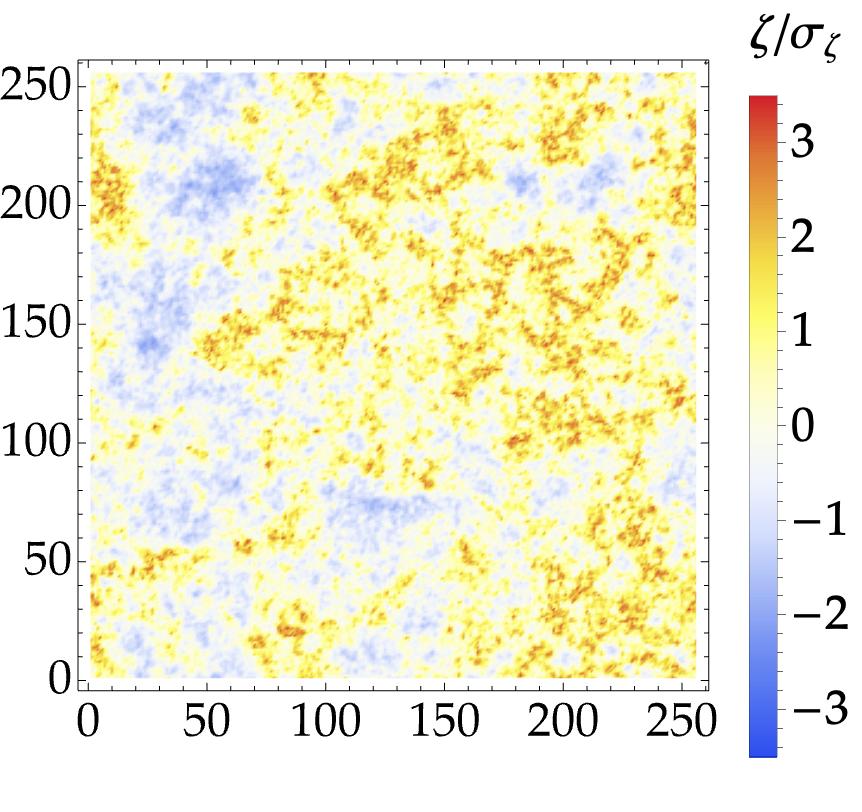}
            \subcaption*{``Quadratic" $n=3$}
        \end{minipage} \\\\\\
        \begin{minipage}{0.3\hsize}
            \centering
            \includegraphics[width=0.95\hsize]{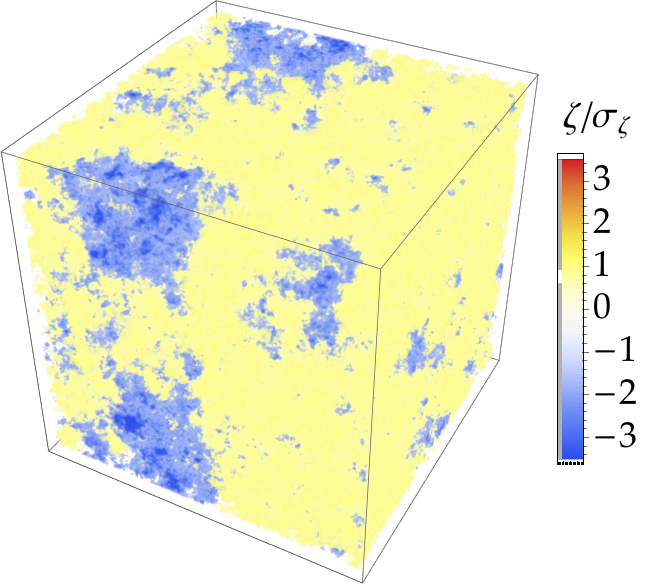}
            \includegraphics[width=0.95\hsize]{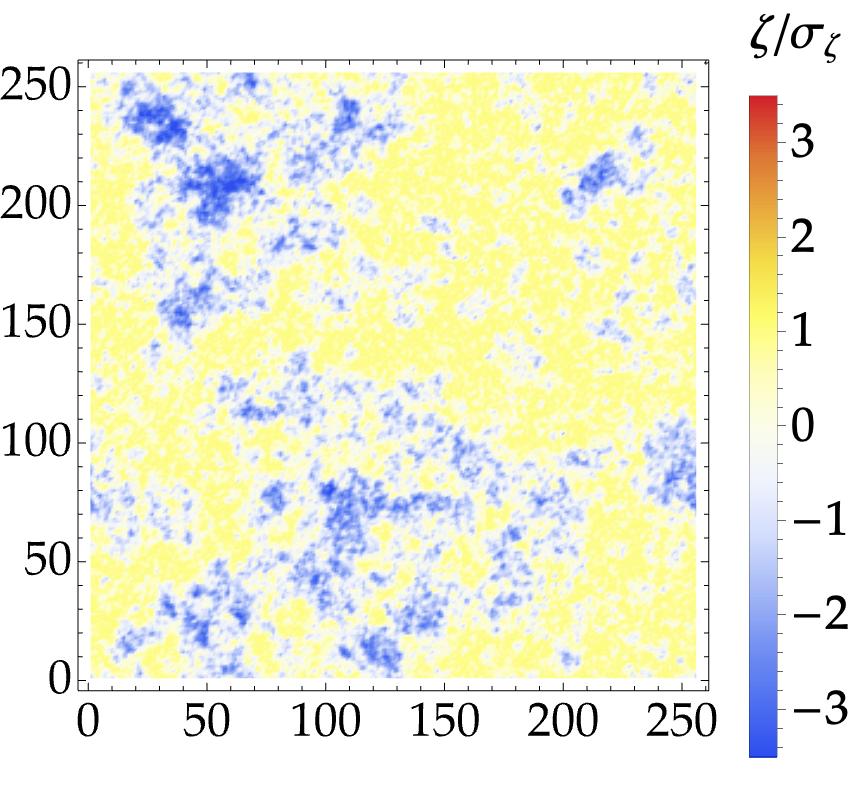}
            \subcaption*{``Cubic" $n=1$}
        \end{minipage} &
        \begin{minipage}{0.3\hsize}
            \centering
            \includegraphics[width=0.95\hsize]{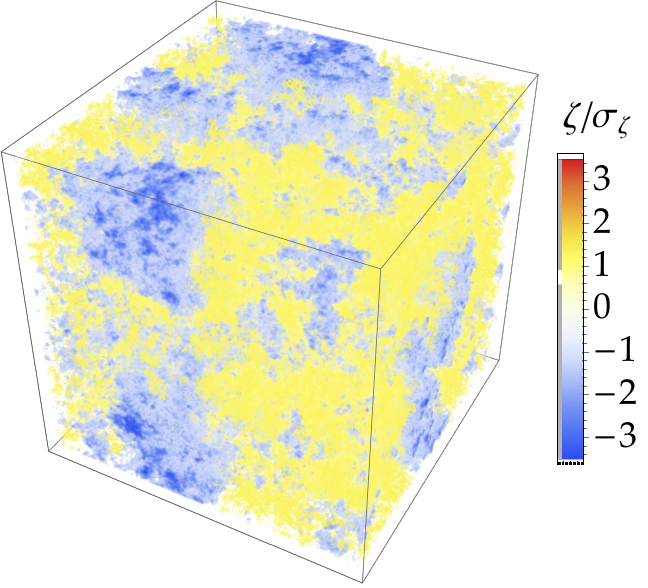}
            \includegraphics[width=0.95\hsize]{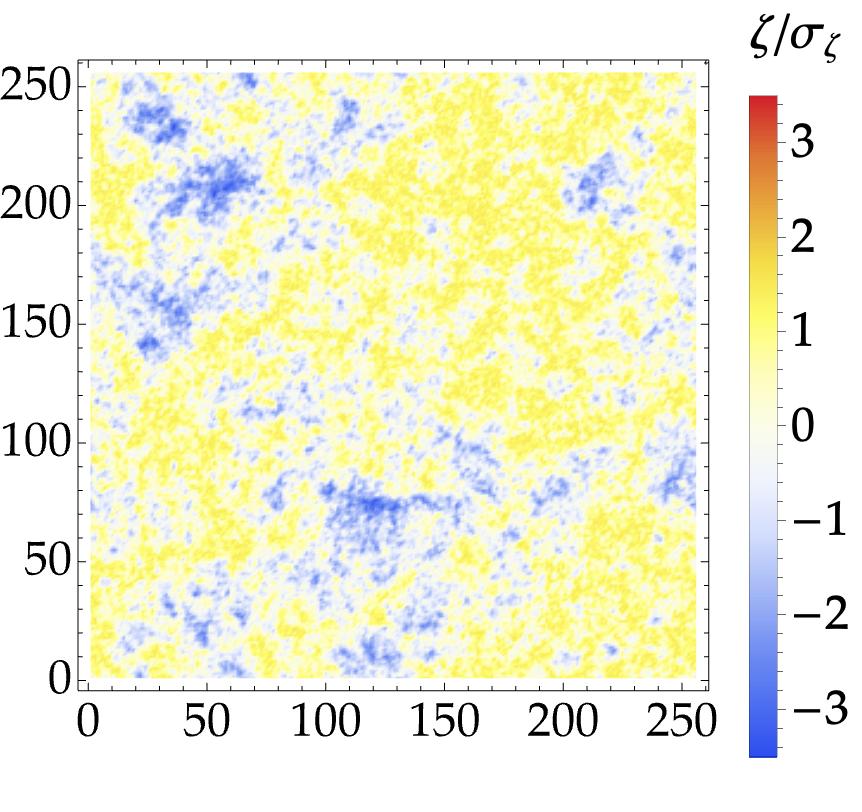}
            \subcaption*{``Cubic" $n=2$}
        \end{minipage} &
        \begin{minipage}{0.3\hsize}
            \centering
            \includegraphics[width=0.95\hsize]{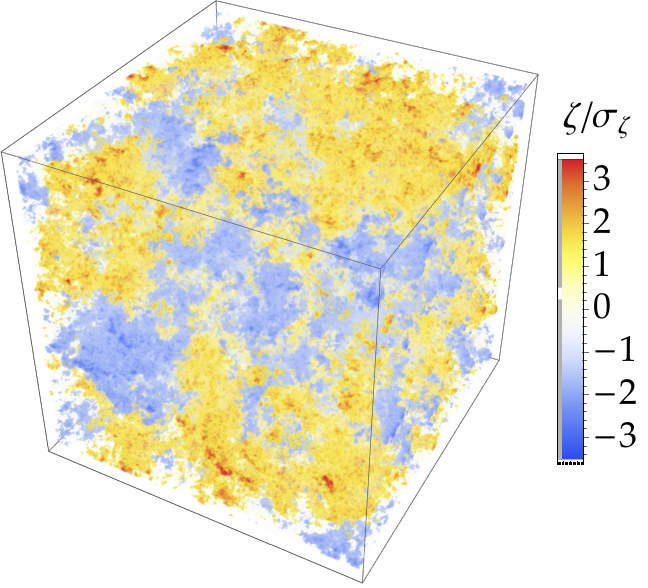}
            \includegraphics[width=0.95\hsize]{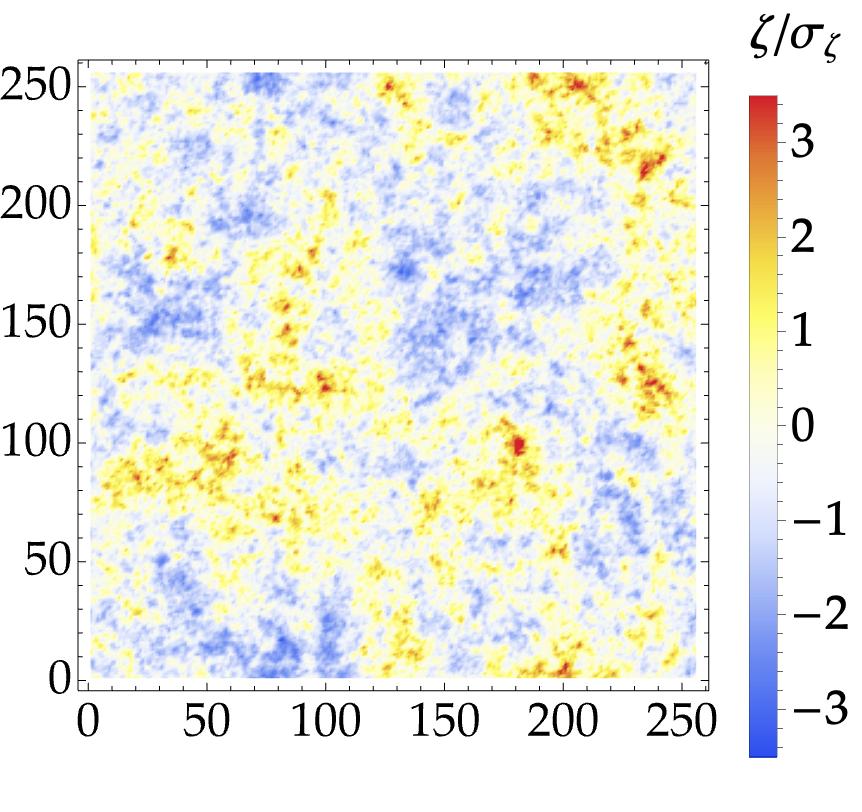}
            \subcaption*{``Quadratic" $n=15$}
        \end{minipage}
    \end{tabular}
    \caption{
    Sample 3D density plots of the curvature perturbation normalised by the standard deviation $\sigma_{\zeta}=\sqrt{\overline{\delta\calN^2}}$ over the simulation box in \ac{STOLAS} and its 2D slice at $z=128$.
    From top-left to bottom-right, they are ``Quadratic'' $n=1$, the ``Quadratic'' $n=2$, the ``Quadratic'' $n=3$, the ``Cubic'' $n=1$, the ``Cubic" $n=2$, and the ``Quadratic'' $n=15$.
    }
    \label{fig: maps}
\end{figure}

In this section, we investigate the statistical properties of the curvature perturbation in \ac{STOLAS}.
We consider the six cases, such as ``Quadratic'' $n=1$, $2$, $3$, and $15,$ and ``Cubic'' $n=1$ and $2$.
The parameters are summarised in Table~\ref{tab: parameters}, which is the same as Ref.~\cite{Murata:2025onc}.
The initial field value of the inflaton is $\phi_\ui=\phi_{\uc}+3/\mu_{1}$, which roughly corresponds to the $3$ e-folds before $\phi_{\uc}$, and others are set to zero.
We adopt the number of grid as $N_{L}=256$.
In Fig.~\ref{fig: maps}, we show the sample 3D density plots of the curvature perturbation (upper panels) and their 2D slice (lower panels) at $z=N_{L}/2=128$ for each model.
The time to generate the noise map $\Delta W_i(N,\bfx)$ used in one simulation was 12 minutes for $N_{L}=256$, and the simulation itself took 8 hours for $n=1$, 11 hours for $n=2$, 13 hours for $n=3$ by Mac Studio with M2 Max (12 core), 64 GB RAM, and 60 hours for $n=15$ with a workstation by HPC systems with Intel Xeon Gold (64 core), 512 GB RAM.

\begin{table}
    \renewcommand{\arraystretch}{1.3}
    \centering    
    \caption{The potential parameters in the ``Quadratic" and ``Cubic" models for any waterfall number $n$. 
    In all cases, we adopt $M=\phi_\uc/\sqrt{2}=0.01\Mpl$.}
    \label{tab: parameters}
    \begin{tabularx}{\hsize}{lCCCCC}
        \toprule
        Model & $\Pi$ & $\Lambda/\Mpl$ & $\mu_{1}/\Mpl$ & $\mu_{2}/\Mpl$ & $\mu_{3}/\Mpl$
        \\
        \hline
        Quadratic 
        & $10$ & $2.66\times 10^{-6}$ & $7.07\times 10^{7}$ & $10.7$ & -
        \\
        Cubic
        & $11$ & $2.41\times 10^{-6}$ & $8.56\times 10^{7}$ & $6.8$ & $0.0375$
        \\
        \bottomrule
    \end{tabularx}
\end{table}

\begin{figure*}
    \!\!\!\!\!\!\!\!\begin{tabular}{c}
        \begin{minipage}{0.33\hsize}
            \centering
            \includegraphics[width=0.95\hsize]{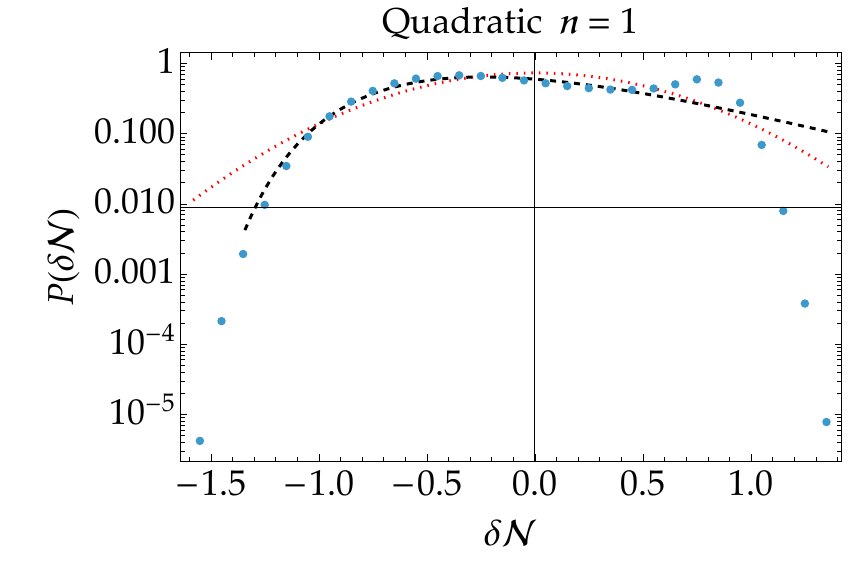}
            \includegraphics[width=0.95\hsize]{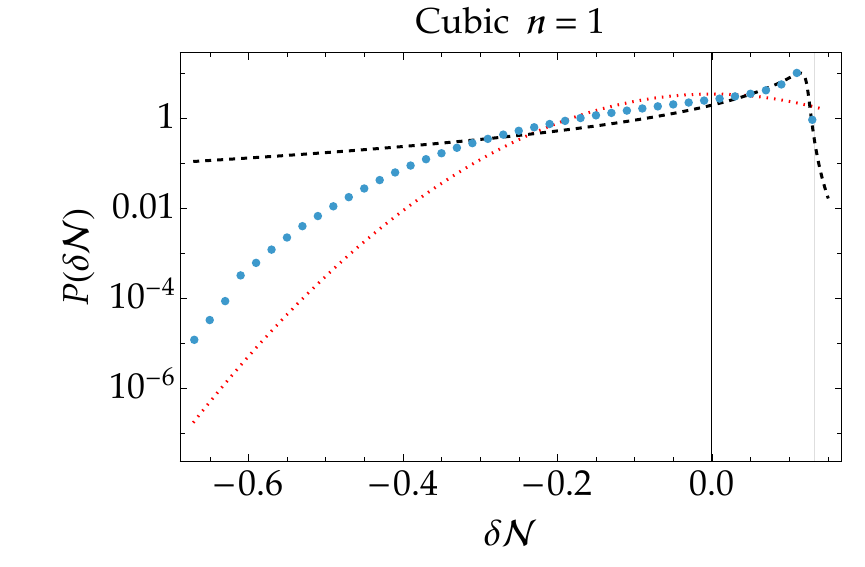}
        \end{minipage}
        \begin{minipage}{0.33\hsize}
            \centering
            \includegraphics[width=0.95\hsize]{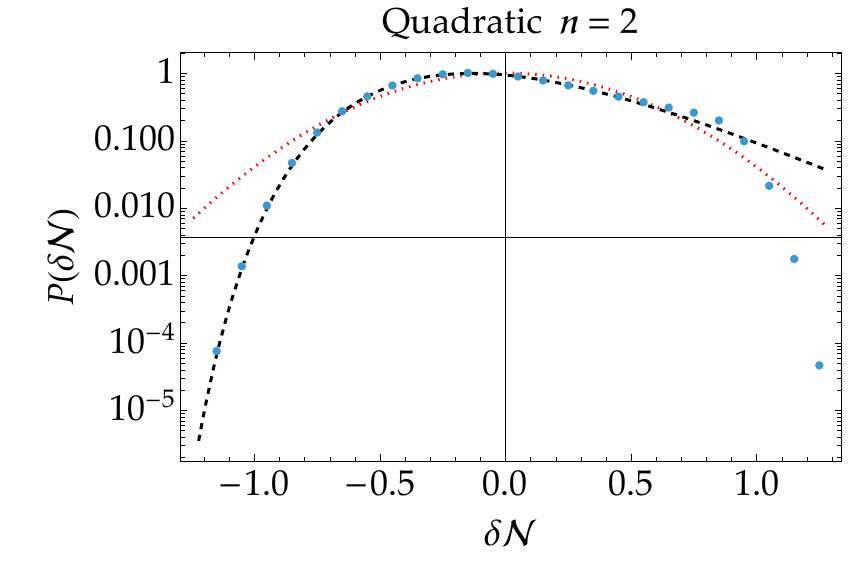}
            \includegraphics[width=0.95\hsize]{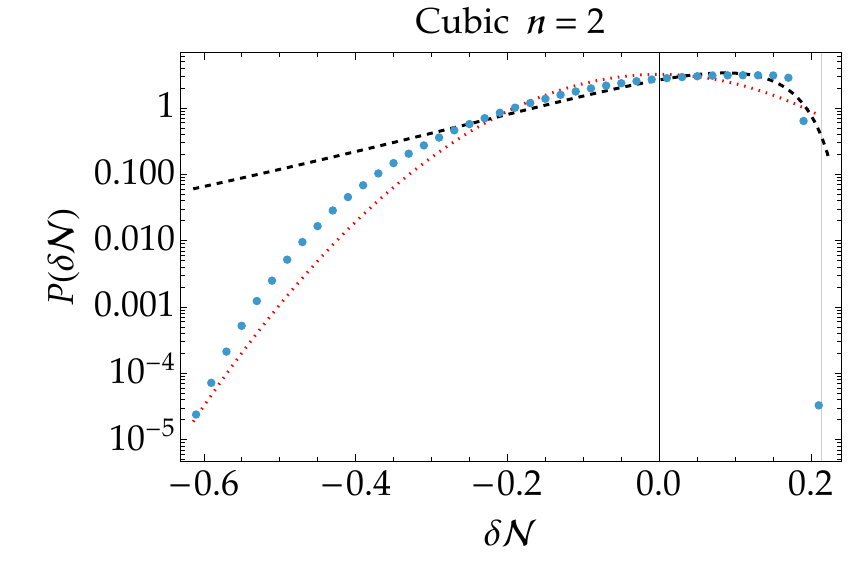}
        \end{minipage}
        \begin{minipage}{0.33\hsize}
            \centering
            \includegraphics[width=0.95\hsize]{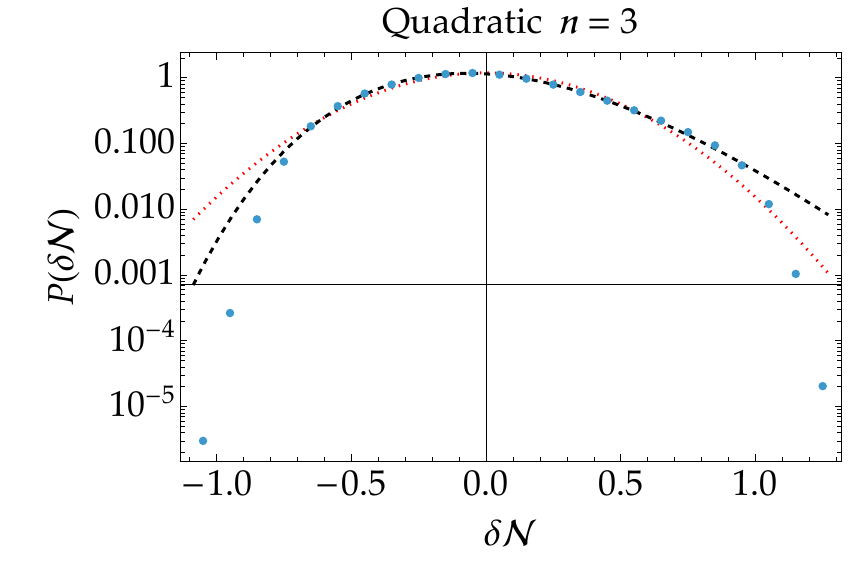}
            \includegraphics[width=0.95\hsize]{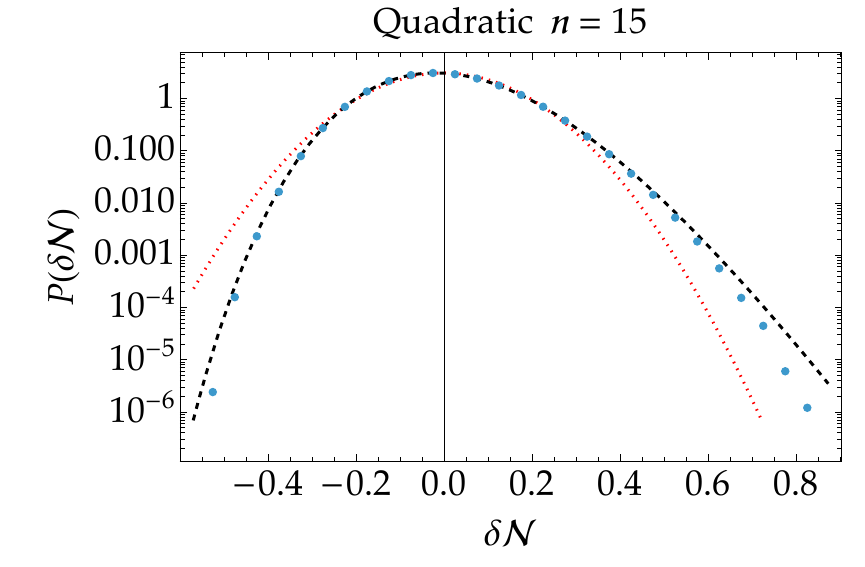}
        \end{minipage}
    \end{tabular}
    \caption{
    \acp{PDF} of the curvature perturbation corresponding to the density plots in Fig.~\ref{fig: maps}.
    The vertical thin line at $\delta\calN=0.133$ for ``Cubic'' $n=1$ and $\delta\calN=0.213$ for ``Cubic'' $n=2$ corresponds to the upper bound on $\delta\calN$.
    }
    \label{fig: pdf}
\end{figure*}

Figure~\ref{fig: pdf} shows the one-point \ac{PDF} of the curvature perturbation, derived by the histograms corresponding to these density plots.
The blue dots represent the simulation results.
The corresponding averages $\overline{\calN}$ and variances $\overline{\delta\calN^2}$ are summarised in Table~\ref{tab: power}.
The black-dashed and the red-dotted lines represent Johnson's $S_U$-distribution fitting and Gaussian fitting, respectively (see Ref.~\cite{Murata:2025onc} for Johnson's $S_U$ fitting).
Needless to say, the Gaussian fitting, and even Johnson's $S_U$-distribution, do not work for $\delta\calN\gtrsim 1$, in contrast to our previous work~\cite{Murata:2025onc}.
The discrepancy between this work and the previous one arises from the averaging procedure discussed in Sec.~\ref{sec: average deltaN}.
That is, the previous work calculated the \ac{PDF} of the curvature perturbation coarse-grained over $k_\sigma^{-1}$ at the end of inflation (i.e., $\zeta$ in Eq.~\eqref{eq: zeta as dN}), while the \ac{PDF} in this work is of the curvature perturbation $\zeta_\uc$ in Eq.~\eqref{eq: zetac} coarse-grained over the lattice grid size, which is much larger than the previous scale.
We have confirmed that the \ac{PDF} similar to that of Ref.~\cite{Murata:2025onc} is reproduced when the averaging procedure is not performed.

One also finds an upper bound $\delta\calN\lesssim0.2$ in the \ac{PDF} for the ``Cubic" models, similarly to Ref.~\cite{Murata:2025onc}.\footnote{Note that the quantitative value of the upper bound for $n=2$ is slightly different from Ref.~\cite{Murata:2025onc}, but it is merely caused by the choice of initial conditions of fields.} 
This is because the extension of $\delta\cal{N}$ by the noise in the waterfall direction has a maximum: the exact hilltop trajectory $\psi_\ur$ supported by a coincidental realisation of the noise. There, inflation is ended purely by the inflaton via the slow-roll violation $\eta_\ur\leq-2$. As the inflaton noise is feeble, it causes an effective upper bound on $\delta\calN$.
This upper bound leads to an interesting conclusion: large positive curvature perturbations such as $\delta\calN\sim 1$ are hardly realised, and hence the \ac{PBH} is expected to seldom form in the ``Cubic'' model.
The upper bound also acts as a \emph{dynamic range compression} in the audio signal processing. 
One finds a bunch of the (yellow) regions with $\zeta$ around the maximum value $\sim 0.2$ almost uniformly in the density plots in Fig.~\ref{fig: maps}, just like ``the sustain of a compressed sound".
Such a unique configuration of the primordial perturbation may affect halo formation, suggesting that $N$-body simulations from these configurations are an interesting direction for future work.

\begin{figure*}
    \!\!\!\!\!\!\!\!\begin{tabular}{c}
        \begin{minipage}{0.33\hsize}
            \centering
            \includegraphics[width=0.95\hsize]{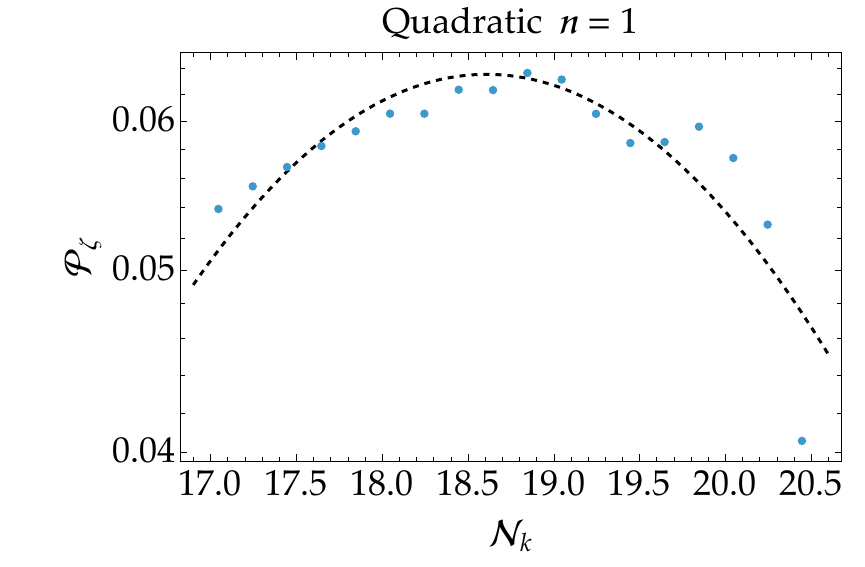}
            \includegraphics[width=0.95\hsize]{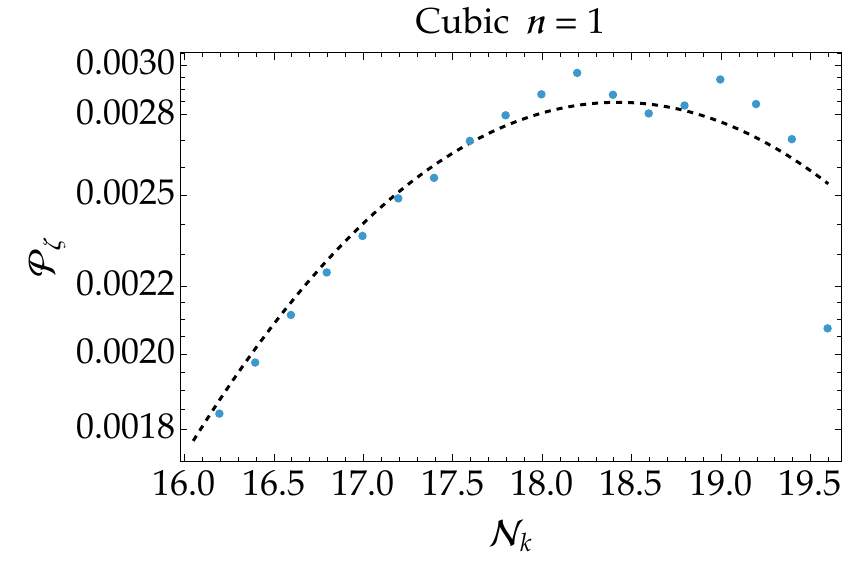}
        \end{minipage}
        \begin{minipage}{0.33\hsize}
            \centering
            \includegraphics[width=0.95\hsize]{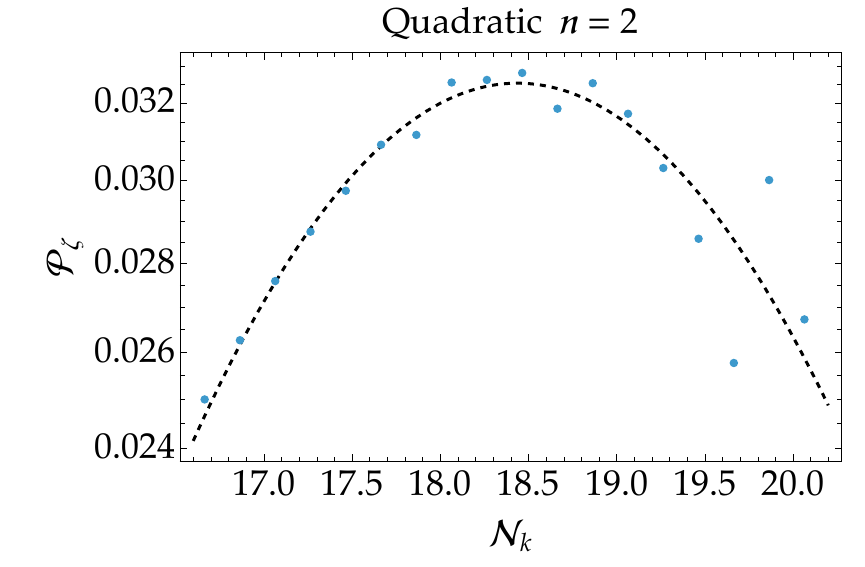}
            \includegraphics[width=0.95\hsize]{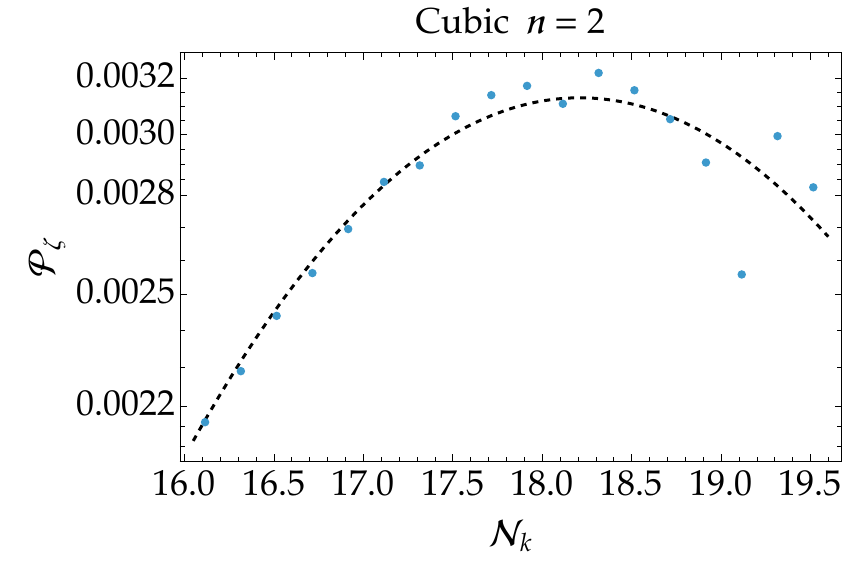}
        \end{minipage}
        \begin{minipage}{0.33\hsize}
            \centering
            \includegraphics[width=0.95\hsize]{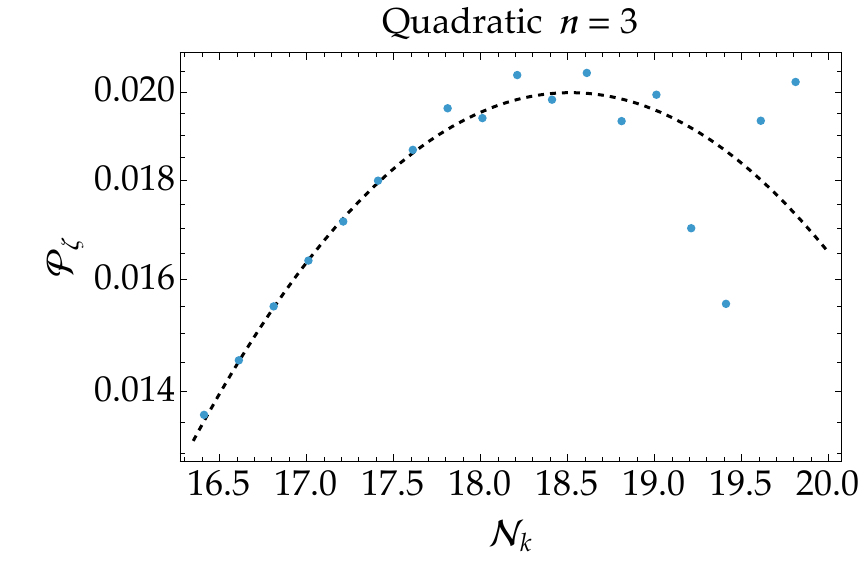}
            \includegraphics[width=0.95\hsize]{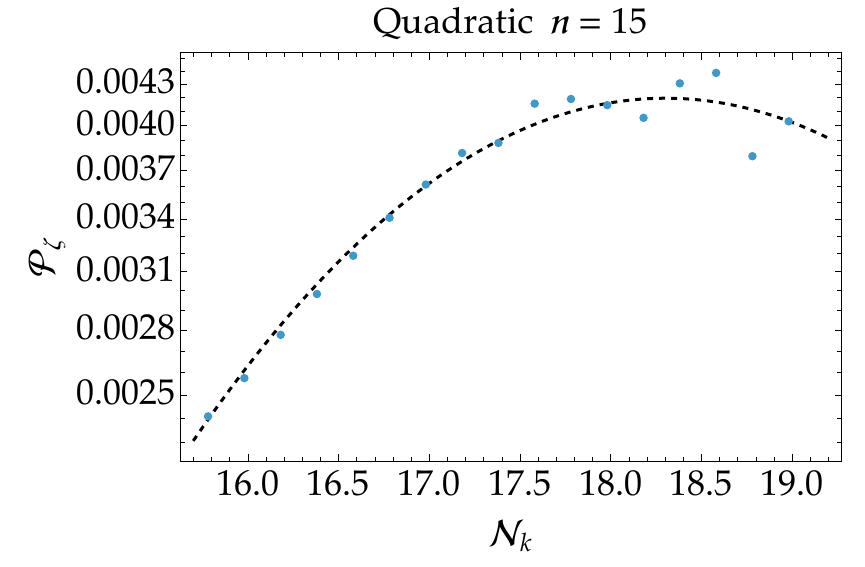}
        \end{minipage}
    \end{tabular}
    \caption{
    Power spectrum for each model as the Fourier transform of the density plot in Fig.~\ref{fig: maps}.
    The blue dots show the result of \ac{STOLAS}.
    The black dashed line shows the analytic formula~\eqref{eq:analytic_formula} with fitting parameters ($N_{\rm water}$, $\calP_{\zeta}^{\rm peak}$), whose values are listed in Table~\ref{tab: power}. We do not show too high $k$ (i.e., high $\calN_k$) modes, which are not reliable due to the discreteness of the lattice.
    }
    \label{fig: power}
\end{figure*}

The power spectrum is immediately obtained by the Fourier transform of the density plots in Fig.~\ref{fig: maps}.\footnote{See the original work~\cite{Mizuguchi:2024kbl} for the detailed algorithm to calculate the continuous power spectrum by the discrete Fourier transform.}
In Fig.~\ref{fig: power}, we show the power spectrum as a function of the backward e-folds $\calN_{k}$, at which $k=k_\sigma$, rather than $k$ itself, to easily compare with the fitting formula~\eqref{eq:analytic_formula}.
The black dashed lines represent the fitting results.
One finds that the fitting formula works well.
The value of $\calP_{\zeta}^{(\rm peak)}$ is roughly proportional to $\propto 1/n$ for the ``Quadratic" models, which is consistent with previous works~\cite{Halpern:2014mca,Tada:2023fvd,Murata:2025onc}.
The ``Cubic" models do not follow this relation because the suppression of the power spectrum in this model is due to the upper bound on $\delta\calN$ determined by the hilltop ($\psi_\ur\sim0$) trajectory and the mean e-folds $\overline{\calN}$, both of which hardly depend on the waterfall number $n$ (see our previous work~\cite{Murata:2025onc} for more details about the upper bound).

Let us comment on the validity of the narrow-sense stochastic-$\delta\calN$ algorithm~\cite{Fujita:2013cna,Fujita:2014tja,Kawasaki:2015ppx}.
Though the success of the fitting formula supports its validity to some extent, one finds slight discrepancies in the values of the fitting parameters between the stochastic-$\delta\calN$ in Ref.~\cite{Tada:2023pue} and the \ac{STOLAS} in this work.
We understand that these discrepancies come from the slow-roll approximation.
Ref.~\cite{Tada:2023pue} uses the slow-roll approximation to neglect the field velocities $\pi_i$ we kept.
Recently, Ref.~\cite{Miyamoto:2025qqm} computed the power spectrum for the ``Quadratic" $n=1$ model using an improved algorithm based on least-squares fitting, within the narrow-sense stochastic-$\delta\calN$ approach.
It keeps $\pi_i$, and the result is quantitatively consistent with ours.
We hence conclude that the narrow-sense stochastic-$\delta\calN$ is validated by \ac{STOLAS}.

\begin{table}
    \renewcommand{\arraystretch}{1.3}
    \centering
    \caption{Simulation results on the average and variance of $\calN$, and the fitting parameters for the analytic formula~\eqref{eq:analytic_formula}.}
    \label{tab: power}
        \begin{tabularx}{\hsize}{lCCCC}
            \toprule
            Model & $\overline{\calN}$ & $\overline{\delta\calN^2}$ & $N_{\rm water}$ & $\calP_{\zeta}^{(\rm peak)}$
            \\
            \hline
            Quadratic $n=1$ & 22.7  & 0.254 & 18.4 & $6.27\times 10^{-2}$
            \\
            Quadratic $n=2$ & 22.2 & 0.156 & 18.4 & $3.25\times 10^{-2}$
            \\
            Quadratic $n=3$ & 21.9 & 0.115 & 18.5 & $2.00\times 10^{-2}$
            \\
            Quadratic $n=15$ & 21.3 & 0.0170 & 18.3 & $4.20\times 10^{-3}$
            \\
            Cubic $n=1$ & 21.7 & 0.0134 & 18.4 & $2.85\times 10^{-3}$
            \\
            Cubic $n=2$ & 21.6 & 0.0156 & 18.2 & $3.13\times 10^{-3}$ \\
            \bottomrule
        \end{tabularx}
\end{table}

\section{Topological defect}%
\label{sec: defect}

Hybrid inflation is known to be accompanied by topological defects in general~\cite{Linde:1993cn}, as the waterfall dynamics spontaneously break the global $\gO(n)$ ($\gZ_2$ for $n=1$) symmetry. 
The models with $n=1$, $n=2$, and $n=3$ lead to the domain wall, cosmic string, and monopole, respectively.\footnote{It is also known that any non-Abelian compact global symmetry is accompanied by the texture if it is completely broken~\cite{Davis:1986nr,Davis:1987yg,Turok:1989ai}, though the $\gO(n-1)$ symmetry remains for $n\geq3$ even after the end of inflation in our models.}
These defects correspond to the $\psi_{\ur}\sim 0$ regions, which are not allowed to roll down toward the potential minimum due to topology.
On the other hand, a small $\psi_\ur$ is na\"ively expected to result in a large $\delta\calN$. In this section, we investigate whether the topological defect could also leave an imprint on the large-scale structure of the curvature perturbation.

\subsection{Snapshots of the waterfall fields}

\begin{figure*}
    \centering
    \begin{tabular}{c}
        \begin{minipage}{0.98\hsize}
            \centering
            \includegraphics[width=0.18\hsize]{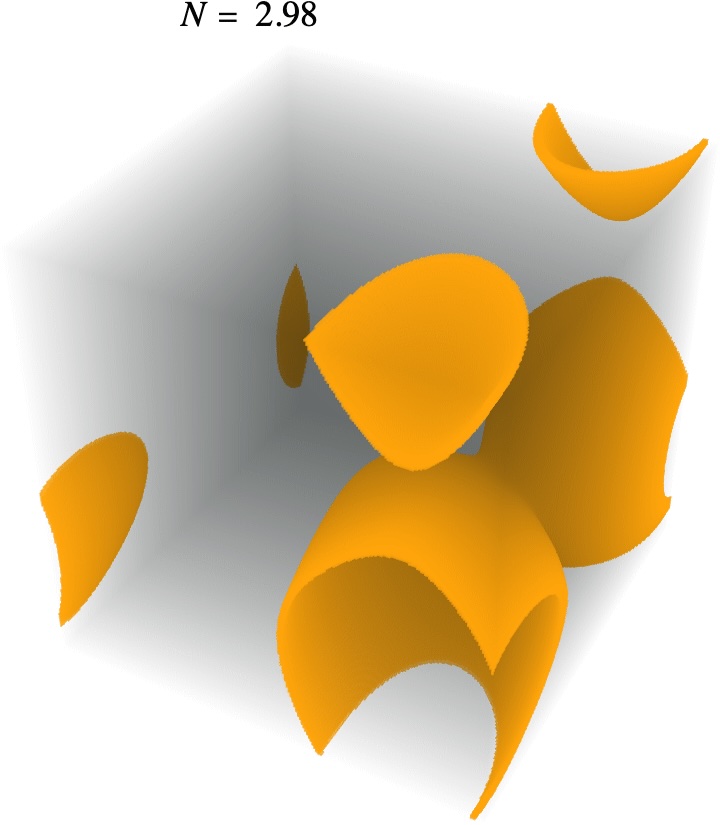}
            \includegraphics[width=0.18\hsize]{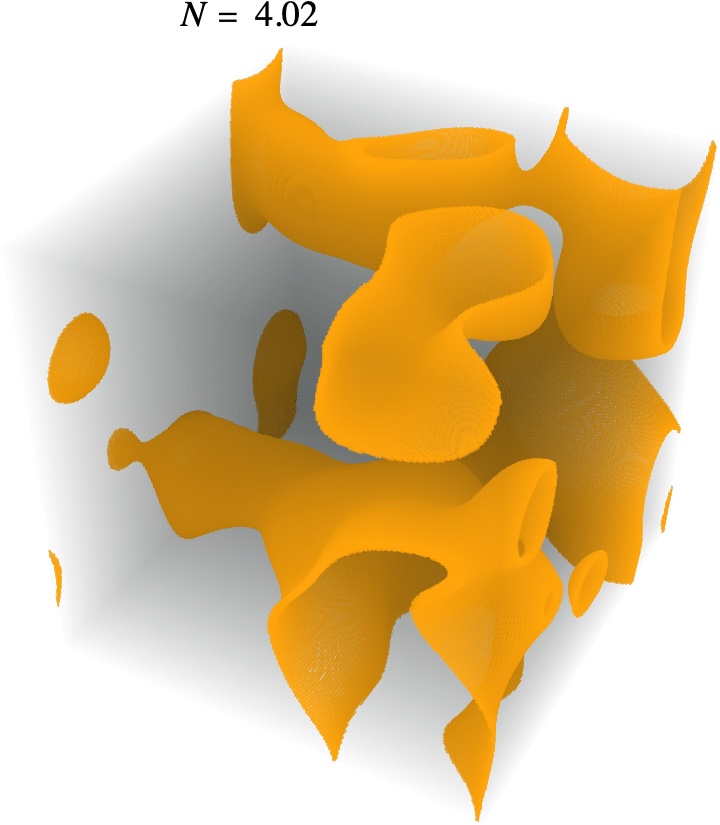}
            \includegraphics[width=0.18\hsize]{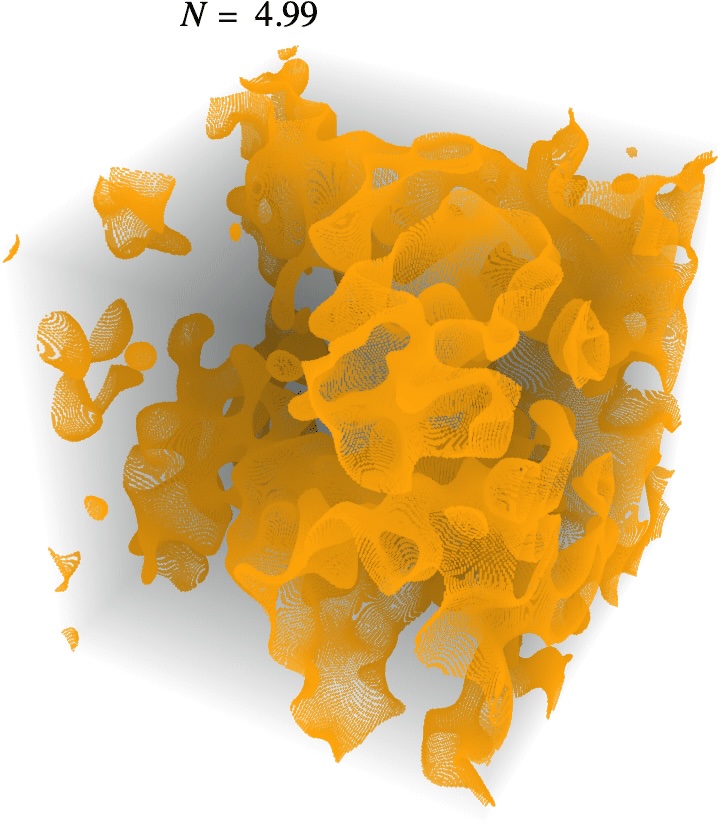}
            \includegraphics[width=0.18\hsize]{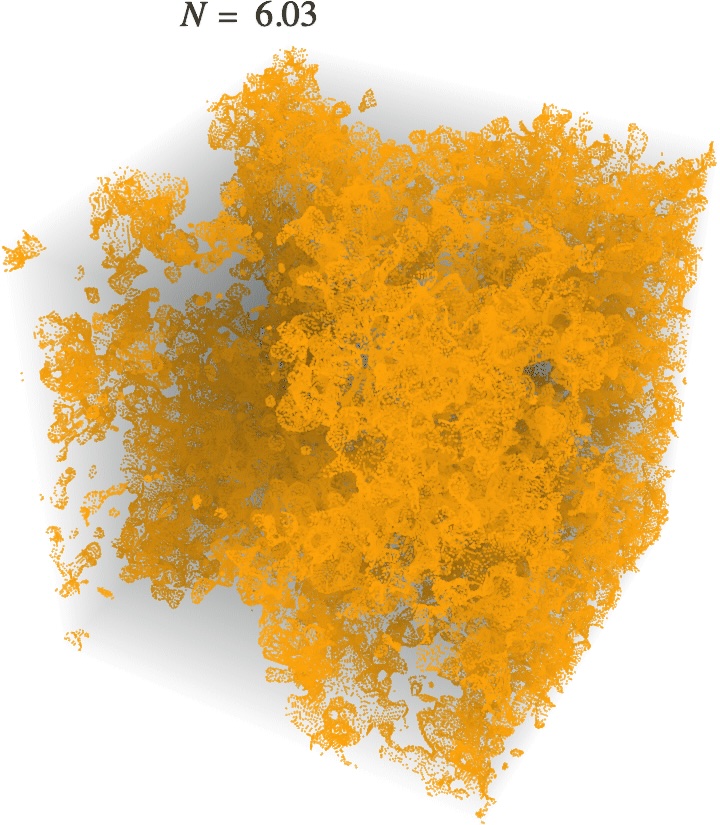}
            \includegraphics[width=0.18\hsize]{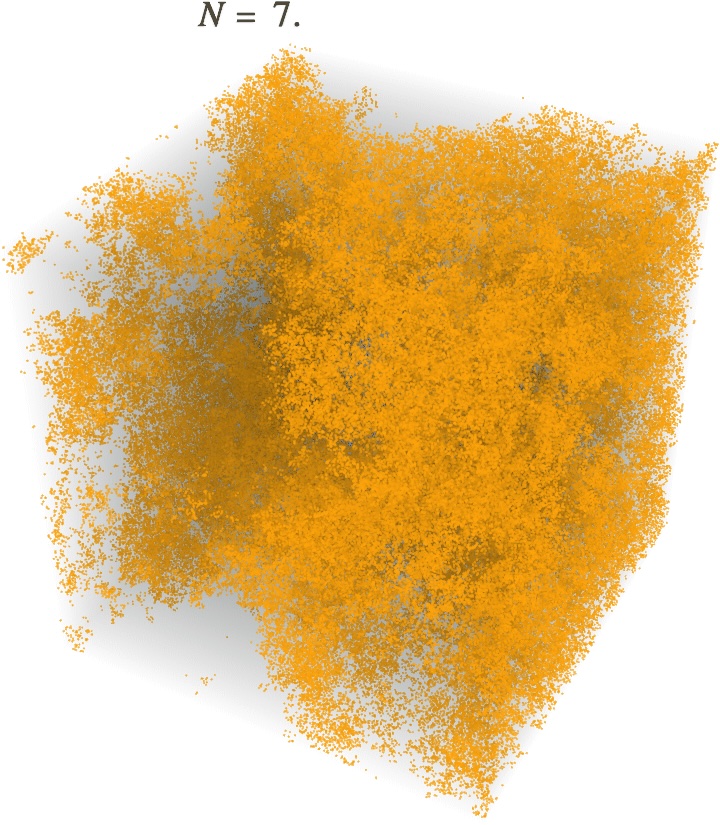}
        \end{minipage}
        \\
        \begin{minipage}{0.98\hsize}
            \centering
            \includegraphics[width=0.18\hsize]{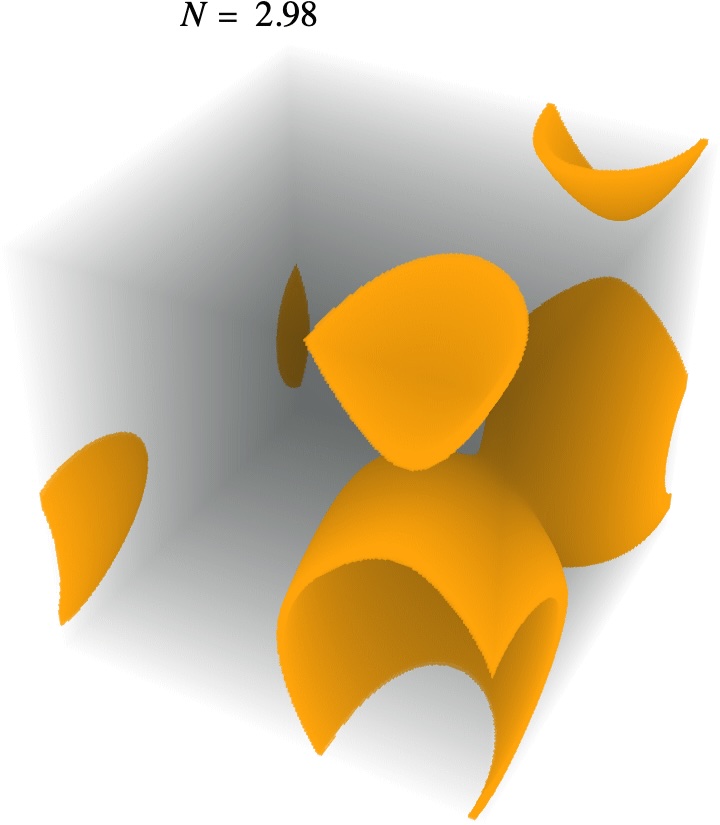}
            \includegraphics[width=0.18\hsize]{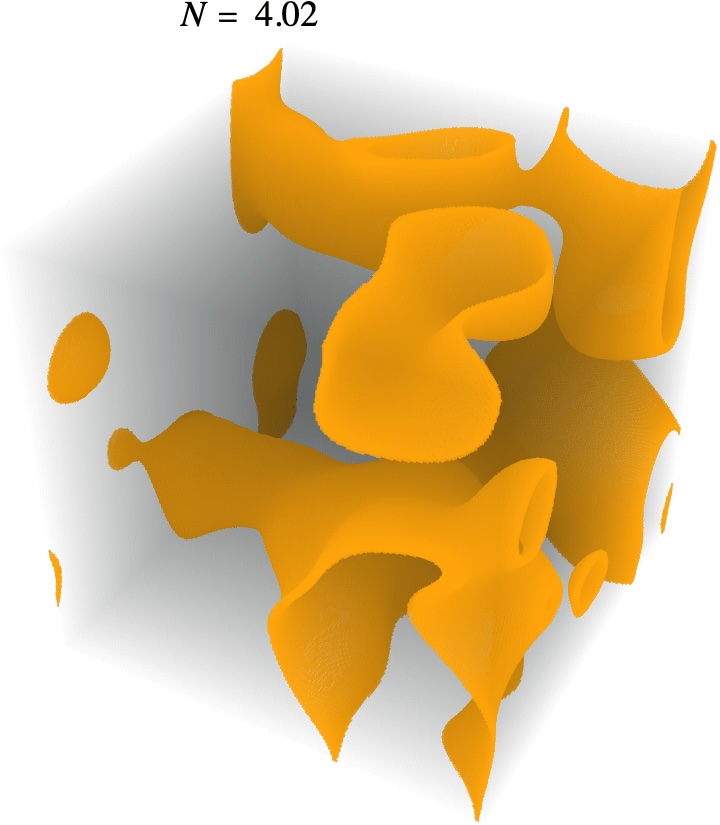}
            \includegraphics[width=0.18\hsize]{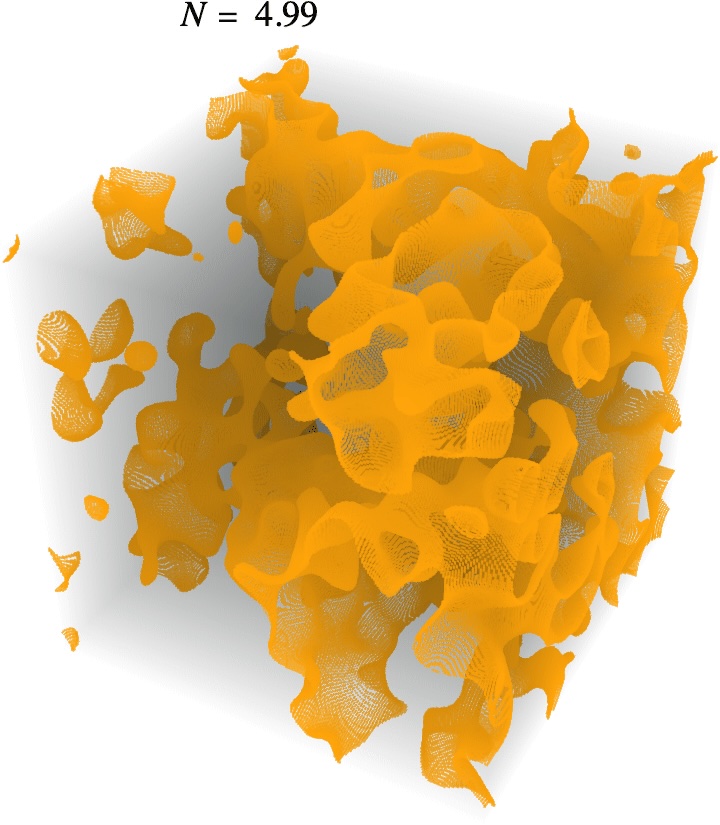}
            \includegraphics[width=0.18\hsize]{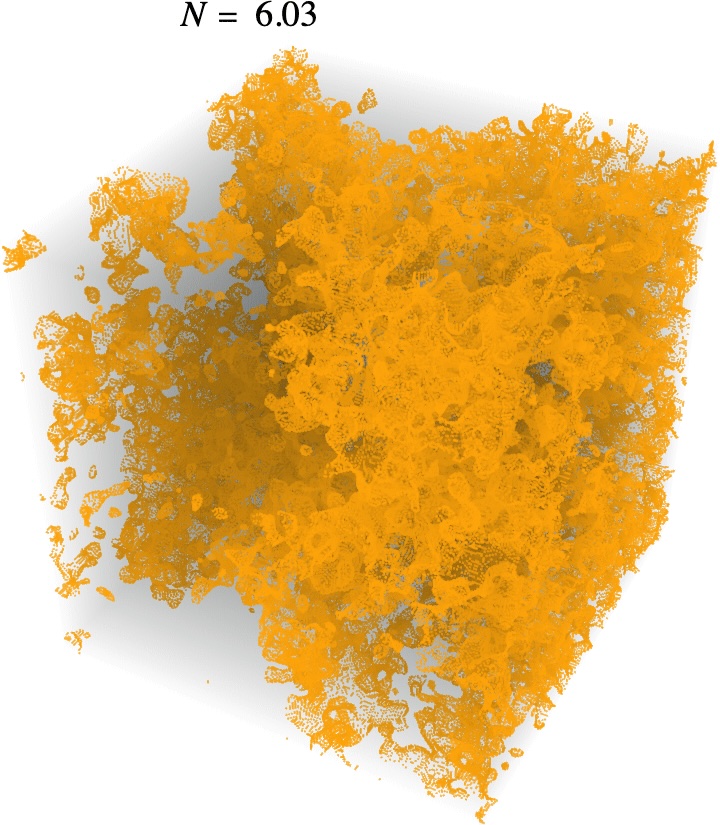}
            \includegraphics[width=0.18\hsize]{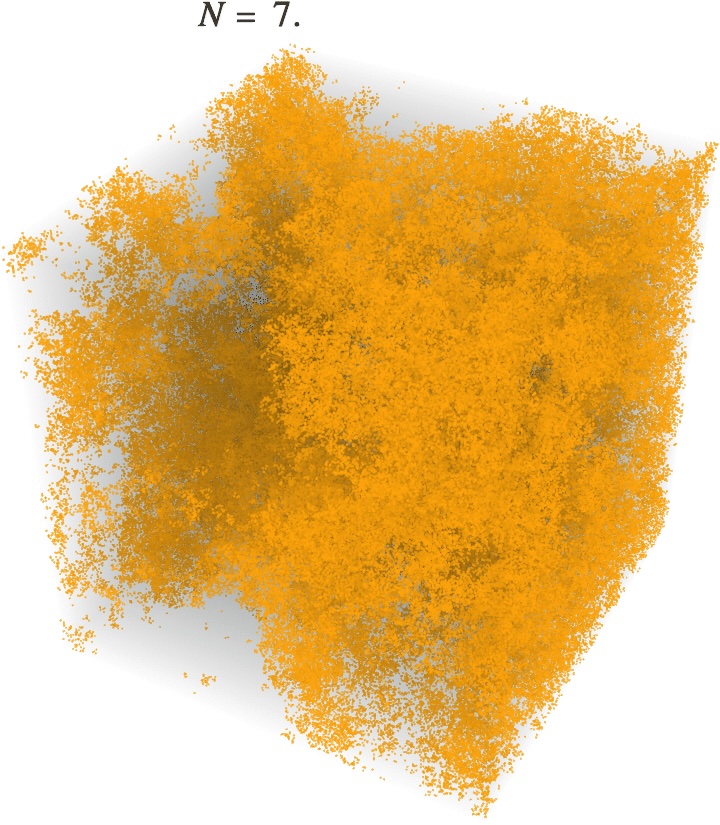}
        \end{minipage}
        \\
        \begin{minipage}{0.98\hsize}
            \centering
            \includegraphics[width=0.18\hsize]{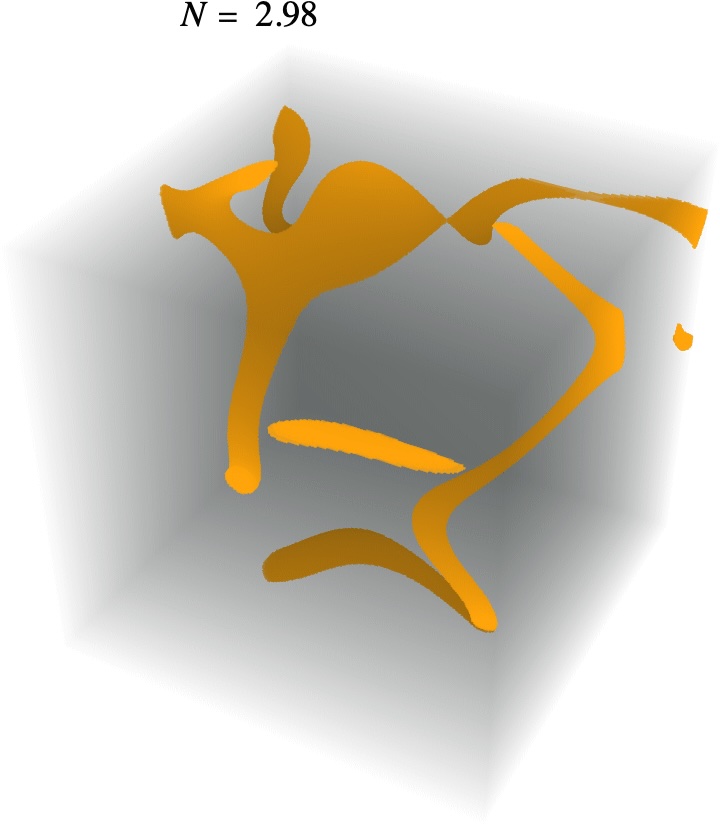}
            \includegraphics[width=0.18\hsize]{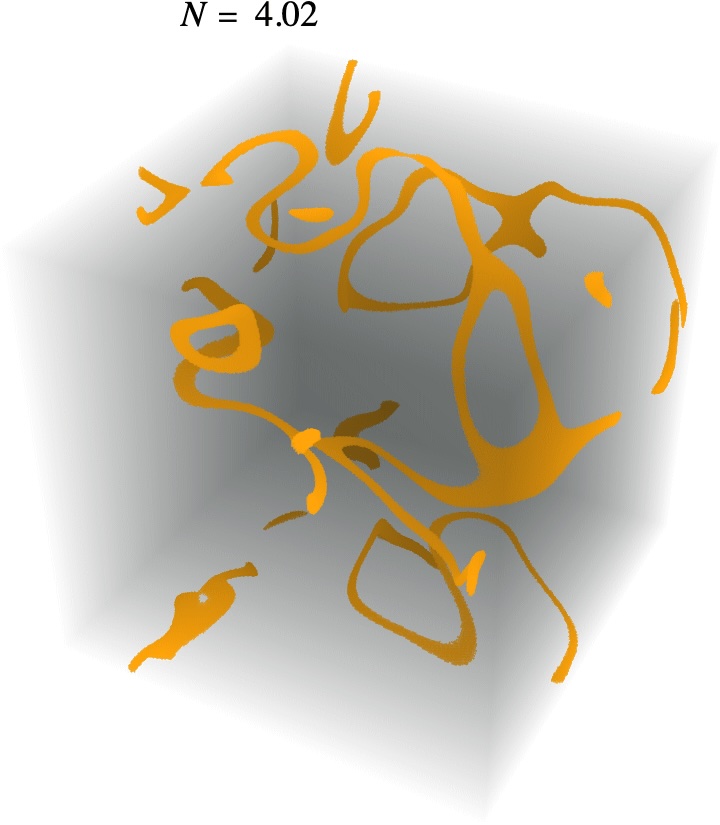}
            \includegraphics[width=0.18\hsize]{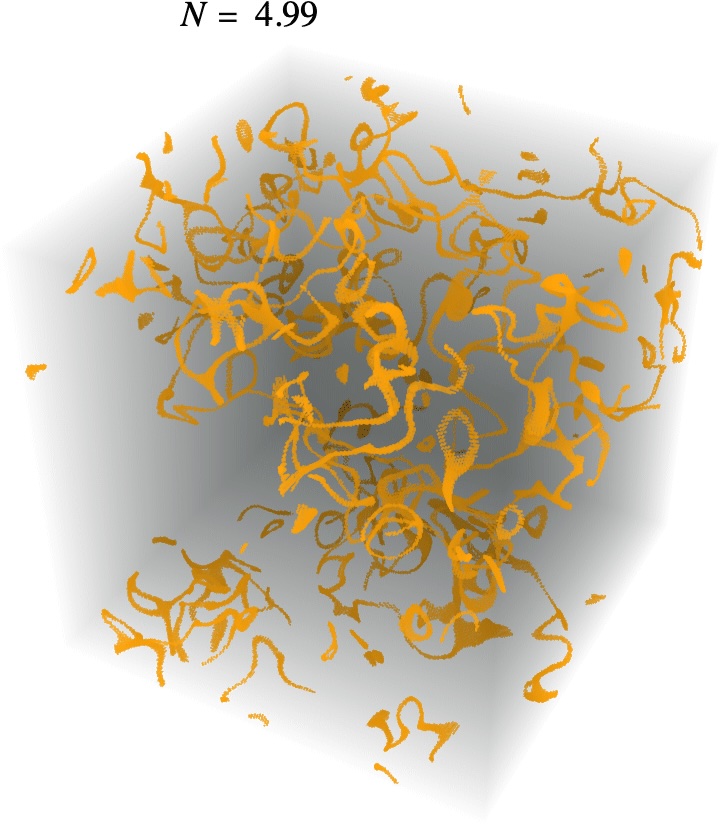}
            \includegraphics[width=0.18\hsize]{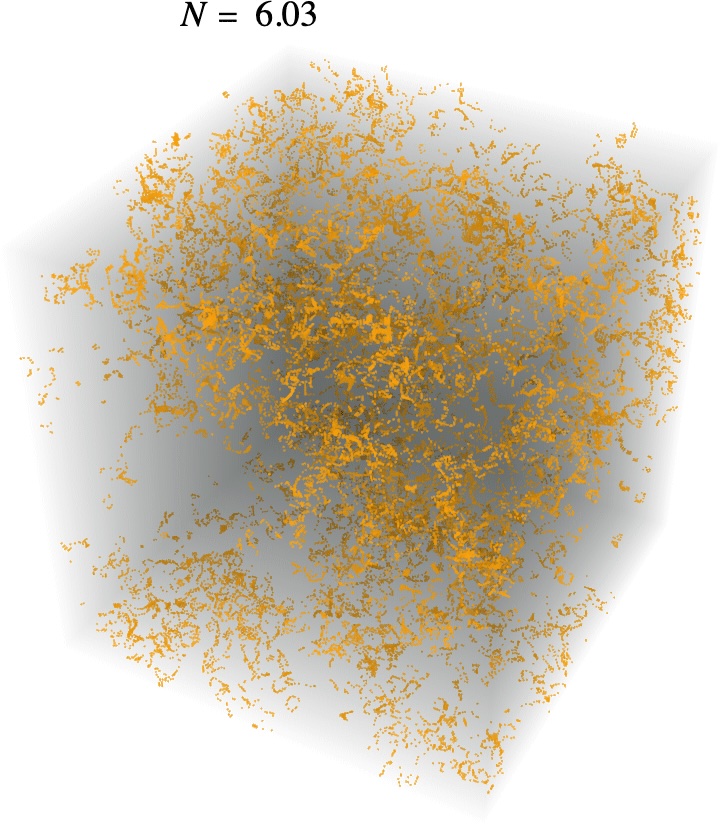}
            \includegraphics[width=0.18\hsize]{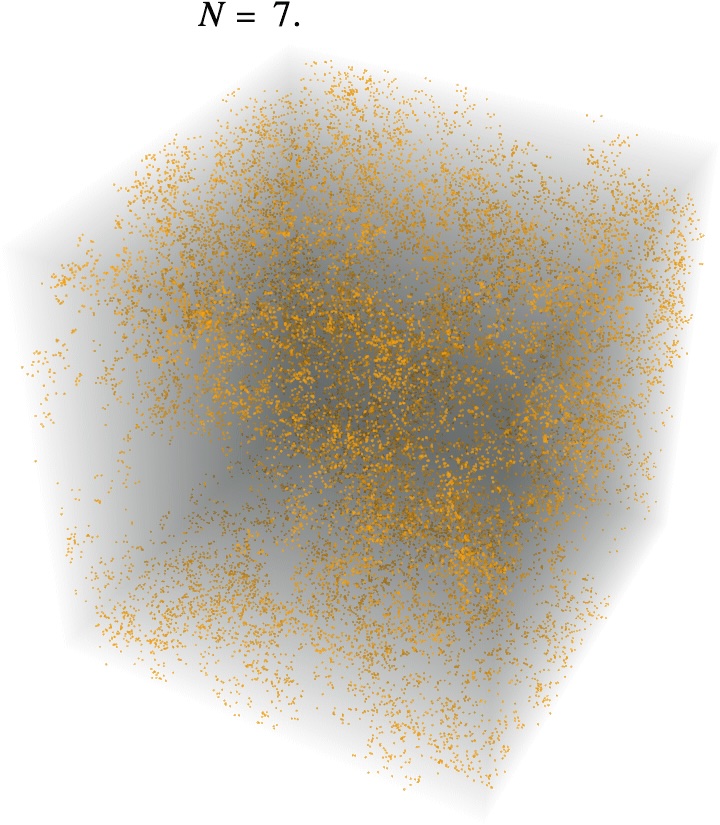}
        \end{minipage}
        \\
        \begin{minipage}{0.98\hsize}
            \centering
            \includegraphics[width=0.18\hsize]{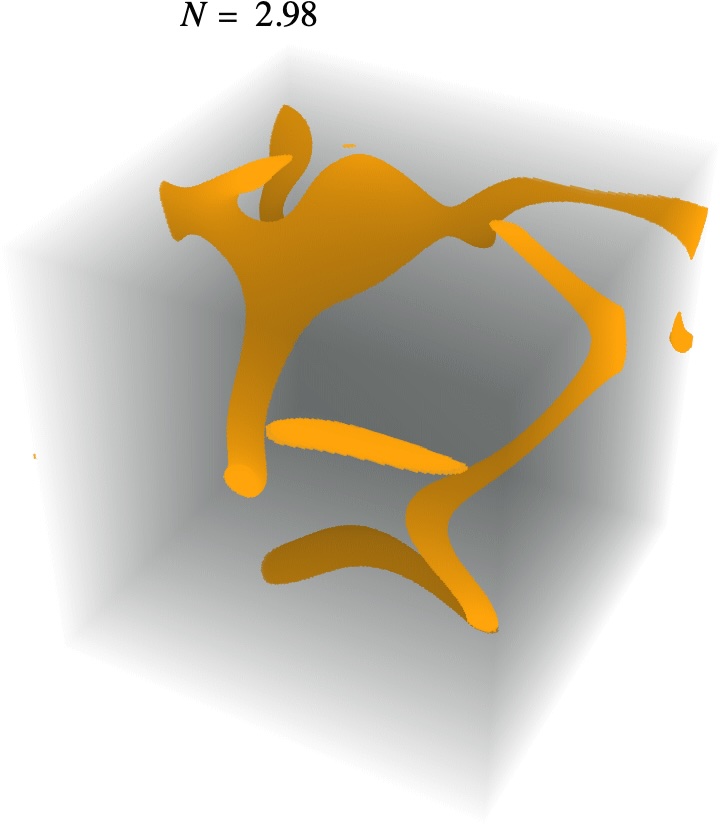}
            \includegraphics[width=0.18\hsize]{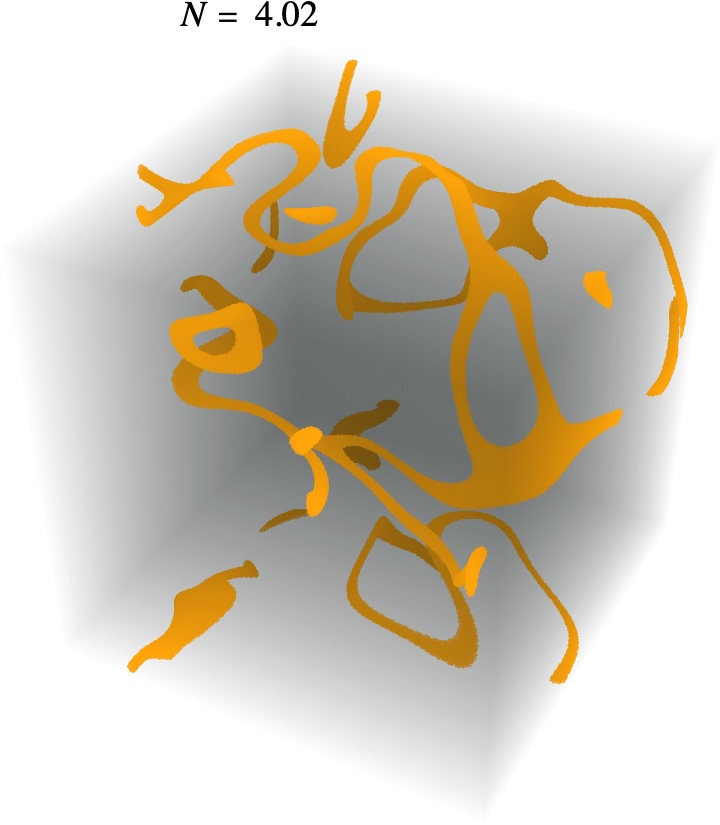}
            \includegraphics[width=0.18\hsize]{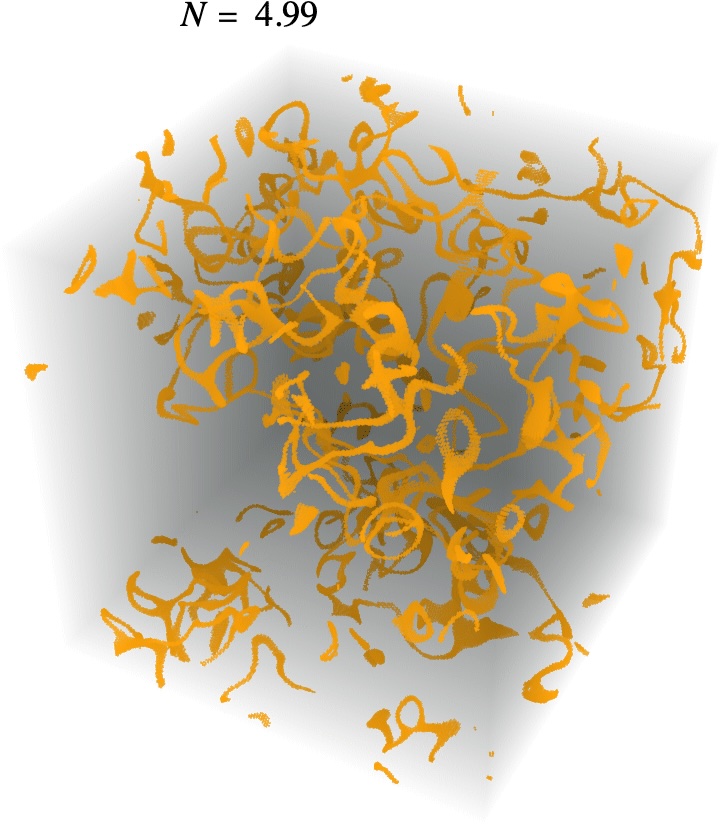}
            \includegraphics[width=0.18\hsize]{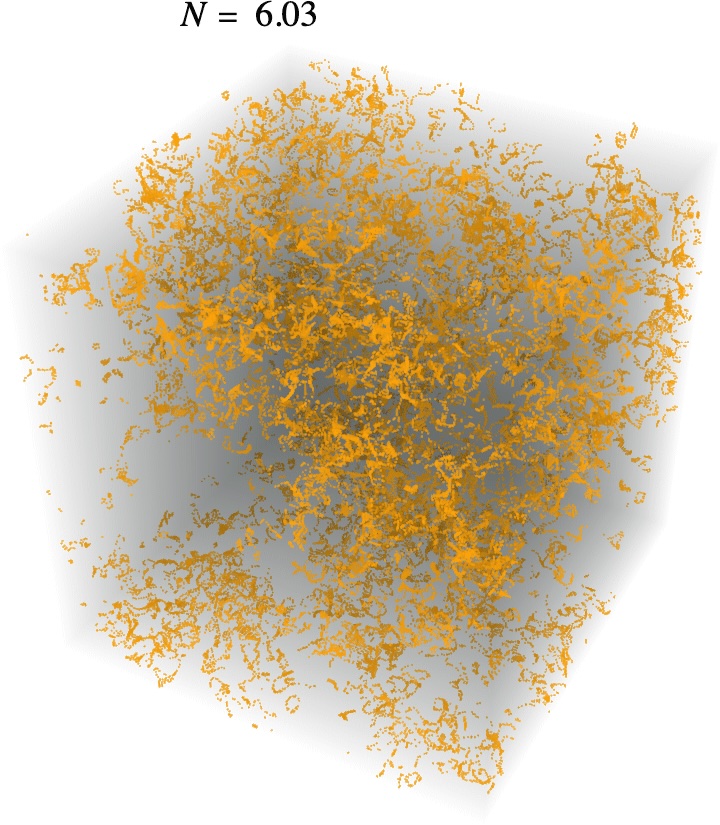}
            \includegraphics[width=0.18\hsize]{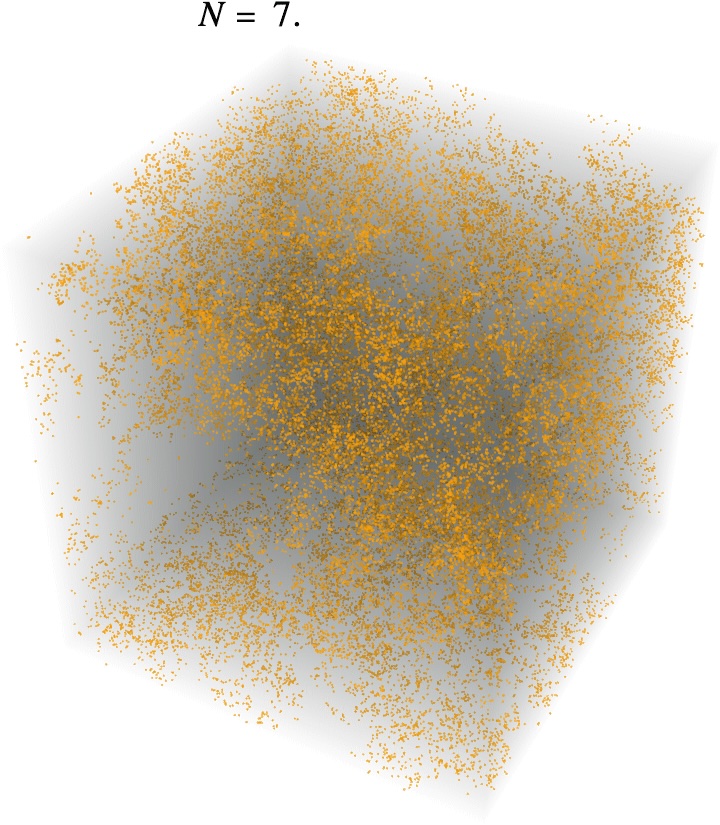}
        \end{minipage}
        \\
        \begin{minipage}{0.98\hsize}
            \centering
            \includegraphics[width=0.18\hsize]{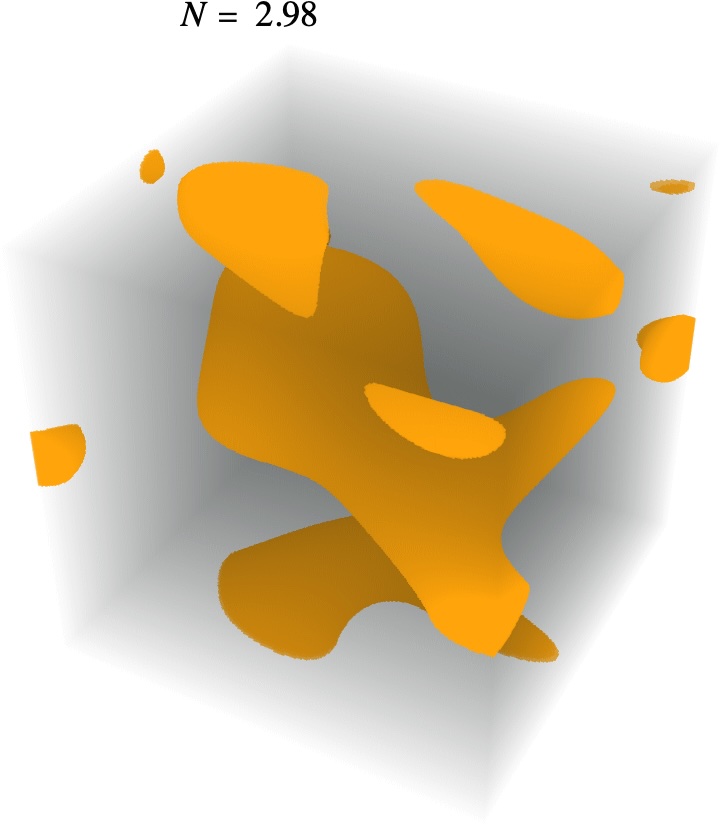}
            \includegraphics[width=0.18\hsize]{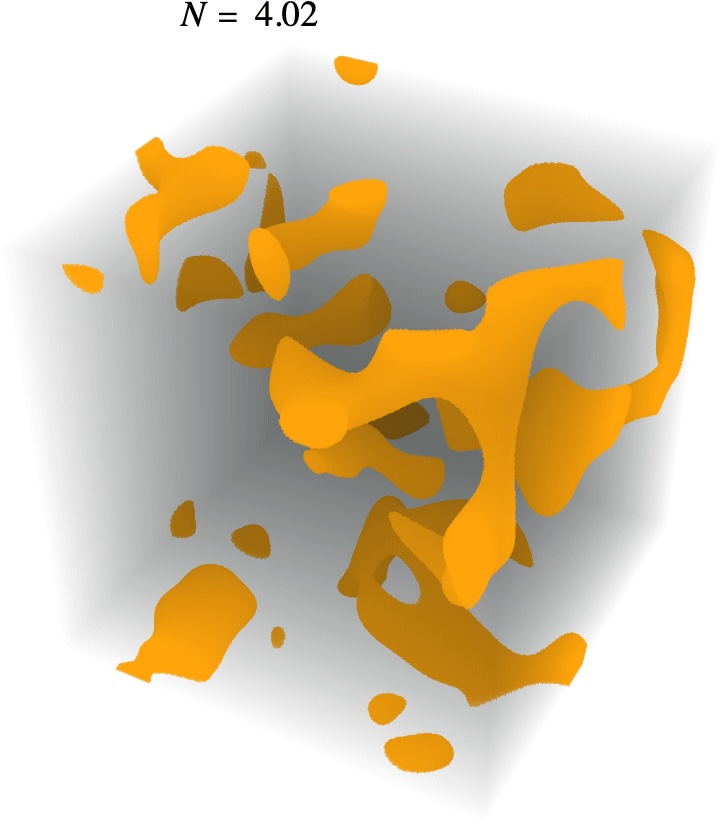}
            \includegraphics[width=0.18\hsize]{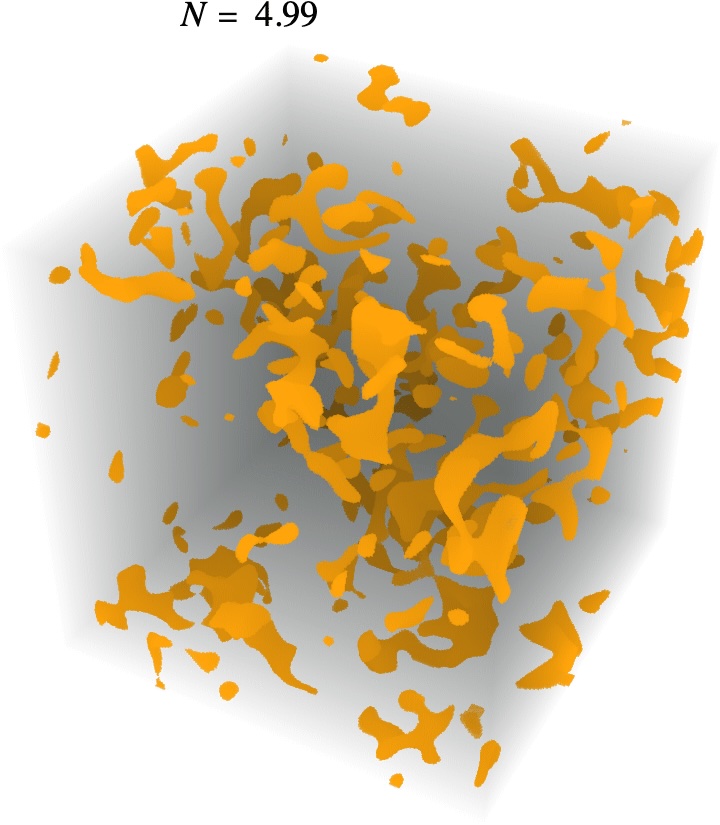}
            \includegraphics[width=0.18\hsize]{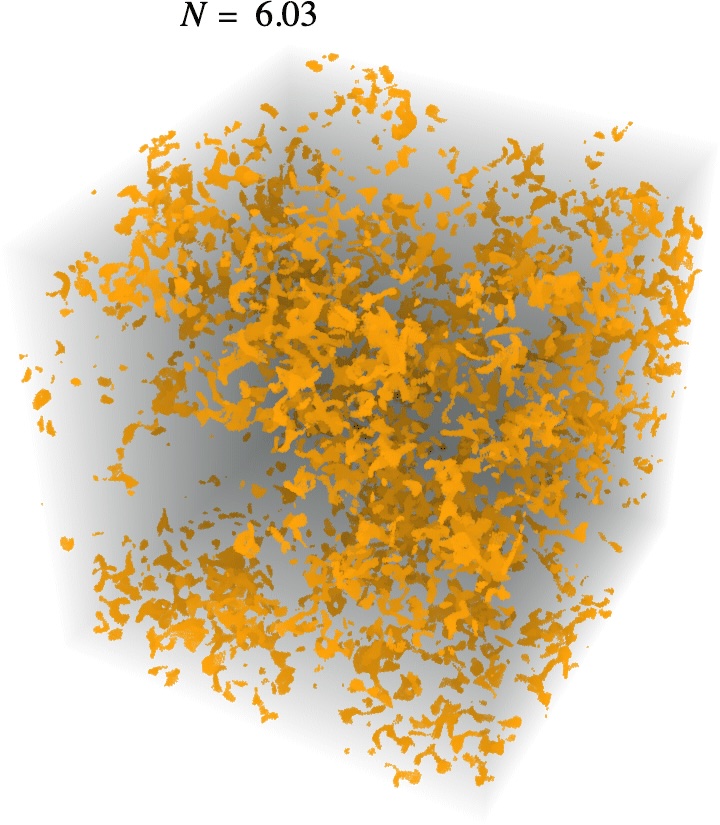}
            \includegraphics[width=0.18\hsize]{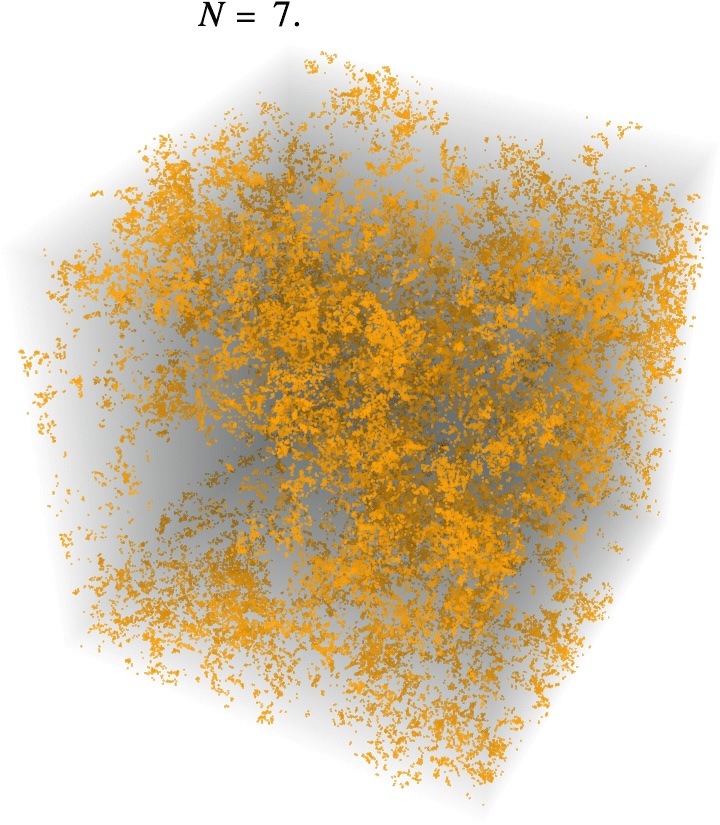}
        \end{minipage}
        \\
        \begin{minipage}{0.98\hsize}
            \centering
            \includegraphics[width=0.18\hsize]{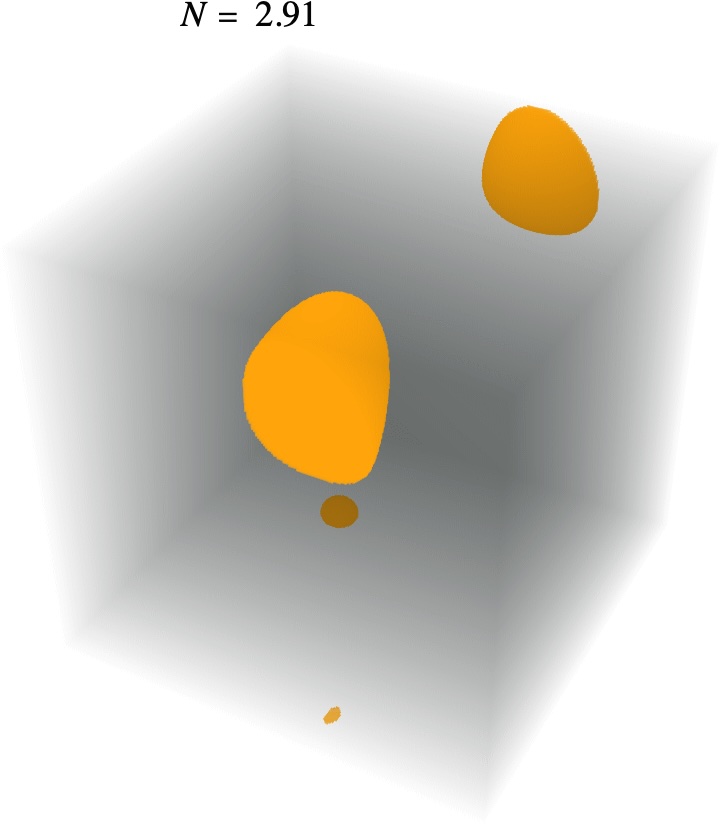}
            \includegraphics[width=0.18\hsize]{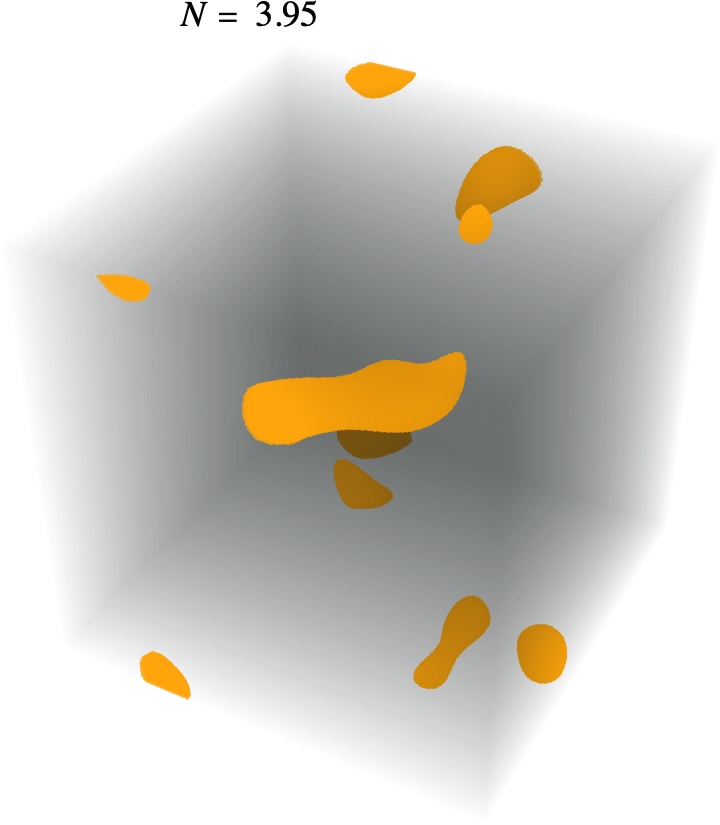}
            \includegraphics[width=0.18\hsize]{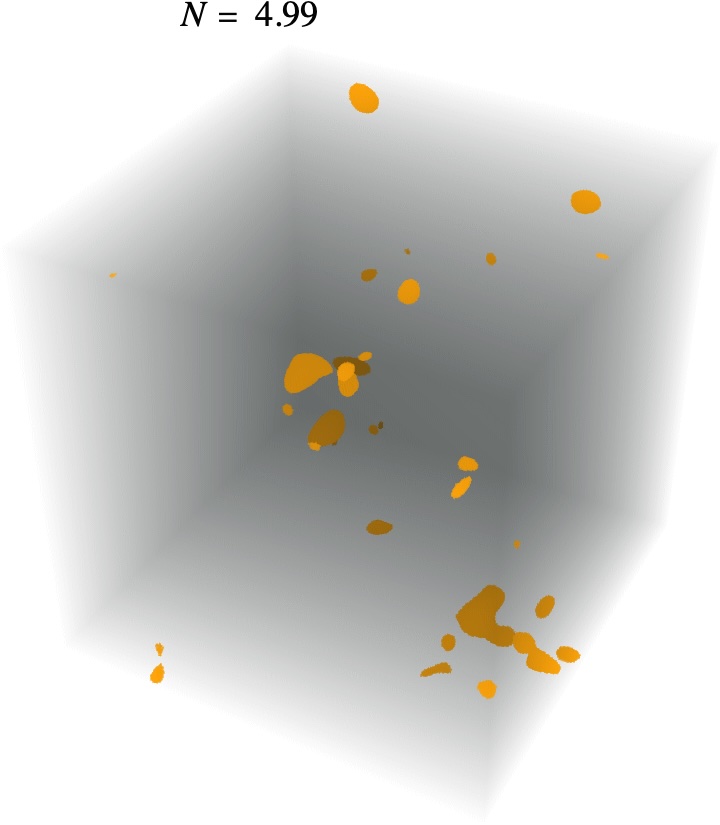}
            \includegraphics[width=0.18\hsize]{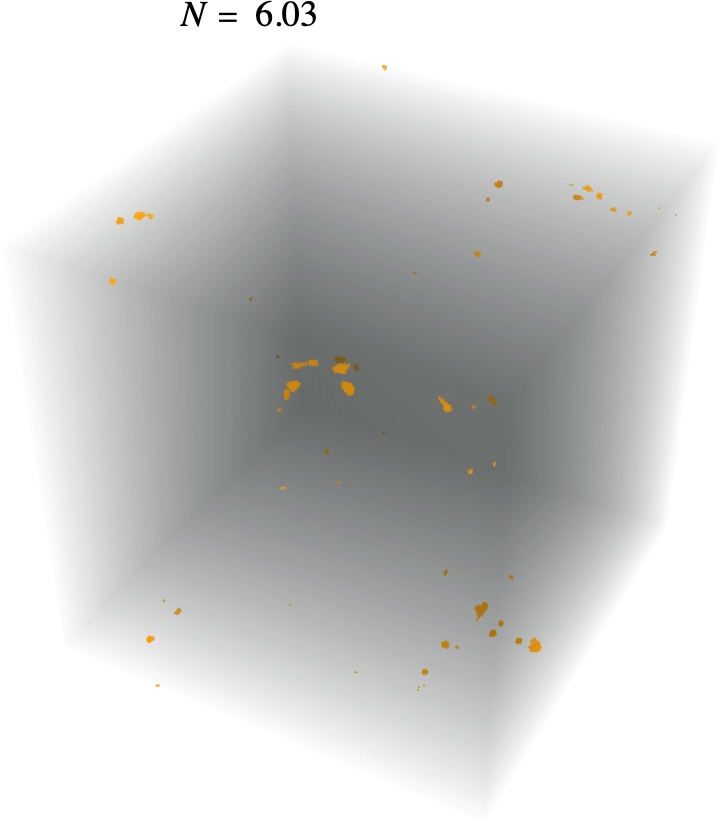}
            \includegraphics[width=0.18\hsize]{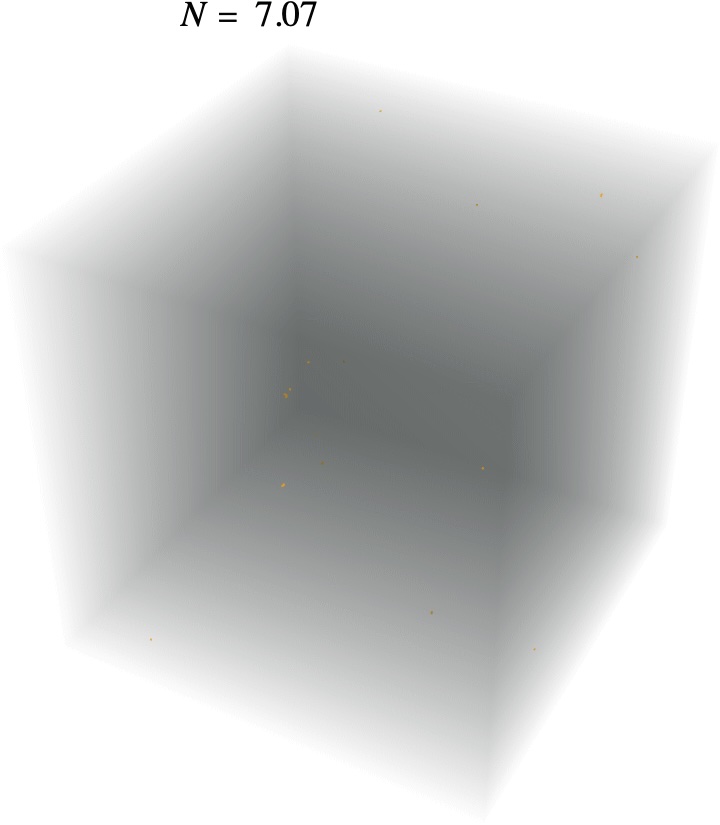}
        \end{minipage}
    \end{tabular}
    \caption{
    Snapshots of the waterfall fields below the threshold values~\eqref{eq: psi threshold}.
    From top row to bottom row, they are ``Quadratic'' $n=1$, the ``Cubic'' $n=1$, the ``Quadratic'' $n=2$, the ``Cubic'' $n=2$, the ``Quadratic'' $n=3$, and the ``Quadratic'' $n=15$.
    Animation of the snapshots can be found \href{https://github.com/STOchasticLAtticeSimulation/STOLAS_dist/tree/main/hybrid-inflation/Fig4_sample}{here}.
    }
    \label{fig: snap}
\end{figure*}

In this subsection, we illustrate the region where the radial direction of the waterfall fields is below the threshold values at each time in Fig.~\ref{fig: snap}.
The threshold values are chosen as
\bae{\label{eq: psi threshold}
    \psi_{\rm r,th} =
    \bce{
        0.05 \sigma_{\psi} & \text{($n=1$; both for ``Quadratic" and ``Cubic"),} \\
        0.1 \sigma_{\psi} & \text{($n=2$; both for ``Quadratic" and ``Cubic"),} \\
        0.5 \sigma_{\psi} & \text{($n=3$),} \\
        2.5 \sigma_{\psi} & \text{($n=15$),}
    }
}
where $\sigma_\psi$ is the typical amplitude of the waterfall fields around the critical point (see Refs.~\cite{Clesse:2015wea,Kawasaki:2015ppx} for its derivation),
\bae{
    \sigma_\psi\coloneqq\frac{\Lambda^2\sqrt{\Pi}}{4\sqrt{3}(2\pi^3)^{1/4}\Mpl}.
}
In particular, for $n=1$ (domain wall), $n=2$ (cosmic string), and $n=3$ (monopole), at around time $N\sim 3\text{--}5$ right after the inflaton passed the critical point $\phi_\uc$, structures emerge that resemble topological defects, which then keep reconnecting due to the stochastic noise into very fine structures.
It has been na\"ively expected that a single or a few defects are formed per the comoving Hubble patch at around the critical point $\phi_\uc$ in the literature (see, e.g., Ref.~\cite{Tada:2024ckk}). 
However, our result suggests that the stochastic reconnection makes their correlation lengths smaller, and a few defects are expected per the comoving Hubble patch at the end of inflation eventually.
The number density and also the spatial configuration of the topological defect in the mild-waterfall models should be investigated in more detail, which we leave for future work.

The monopole-like objects are formed at $N\sim3$ also for $n=15$, but they disappeared by the end of inflation.
It is because they are not topologically stable for $n\geq4$.
One also finds for $n=1$ and $2$ that the ``Quadratic" or ``Cubic" hardly affects the structure because the topology is determined only by the symmetry of the waterfall fields.

\subsection{Euler characteristic of waterfall fields}

We also present the Euler characteristic corresponding to the snapshots exemplified in Fig.~\ref{fig: snap} to quantify the topological features of the waterfall field.
There, the grid points with $\psi_\ur<\psi_{\ur,\uth}$ are understood as cubes with the grid size whose centres are put at the corresponding grids.
In Mathematica, which we used, these cubes are considered a single connected object if they are touching at a vertex, edge, or face.
We referred to the total number of connected objects in the simulation box $\mathscr{N}$.
The Euler characteristic $\chi_{I}$ of the $I$-th object is defined by
\bae{
  \chi_{I} = V_{I} - E_{I} + F_{I}
  \quad
  (I=1,\cdots \mathscr{N}),
}
where $V_{I}$, $E_{I}$, and $F_{I}$ are the number of vertices, edges, and faces of the $I$-th object, respectively.
We compute the total Euler characteristic for each object in the simulation box:
\bae{\label{eq: chi total}
    \chi = \sum_{I=1}^{\mathscr{N}} \chi_{I}.
}
The Euler characteristic of any polyhedron with the surface of a topological sphere (i.e., a single cube or cubes connected only by their faces without a hole in our case) is two.
A single hole reduces it by two, and a connection by edges or vertices increases it by one (see Appendix~\ref{app: EC in mathematica}).
Qualitatively, the Euler characteristic smaller than two on average implies the existence of certain global structures, while two or larger than two suggests more isolated configurations.

\begin{figure*}
    \centering
    \begin{tabular}{c}
        \begin{minipage}{0.32\hsize}
            \centering
            \includegraphics[width=0.95\hsize]{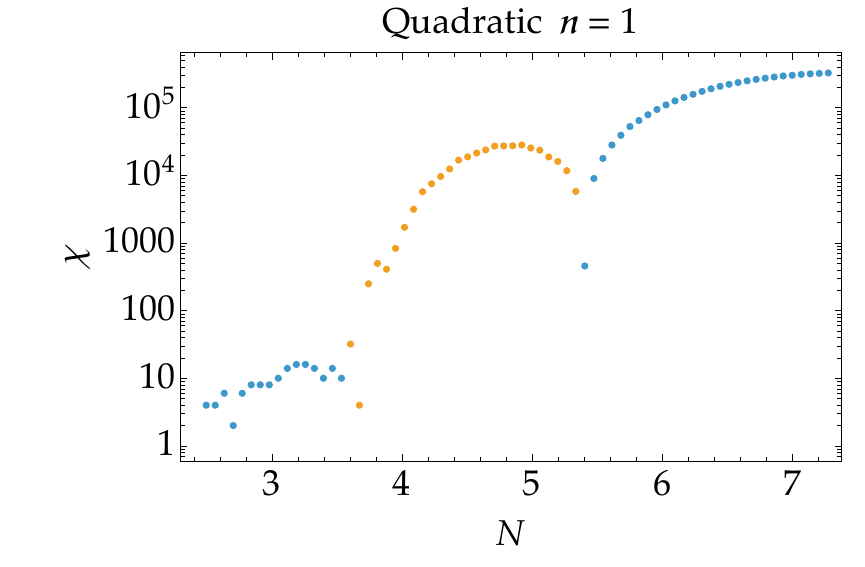}
            \includegraphics[width=0.95\hsize]{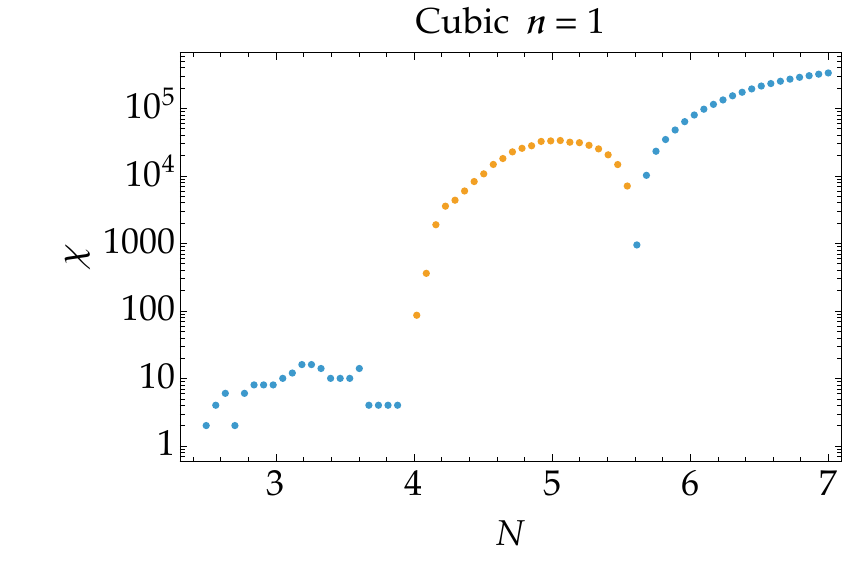}
        \end{minipage}
        \begin{minipage}{0.32\hsize}
            \centering
            \includegraphics[width=0.95\hsize]{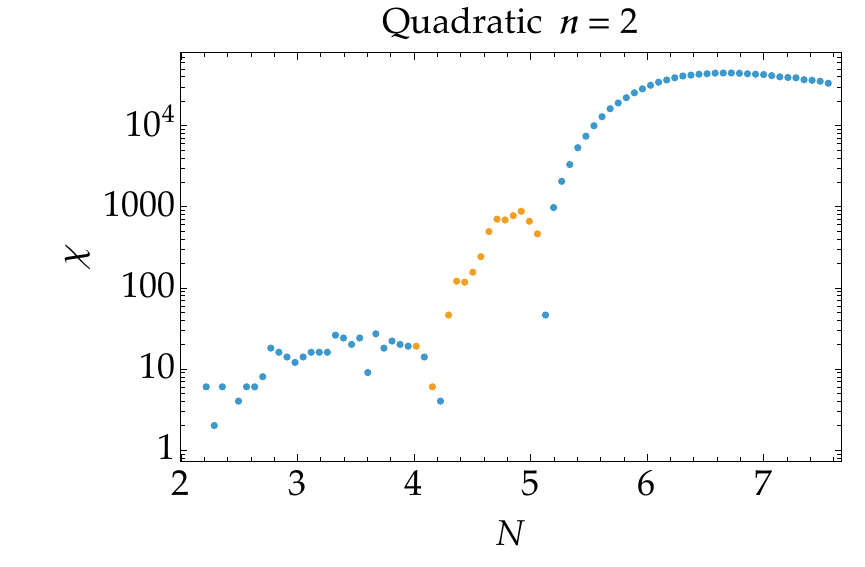}
            \includegraphics[width=0.95\hsize]{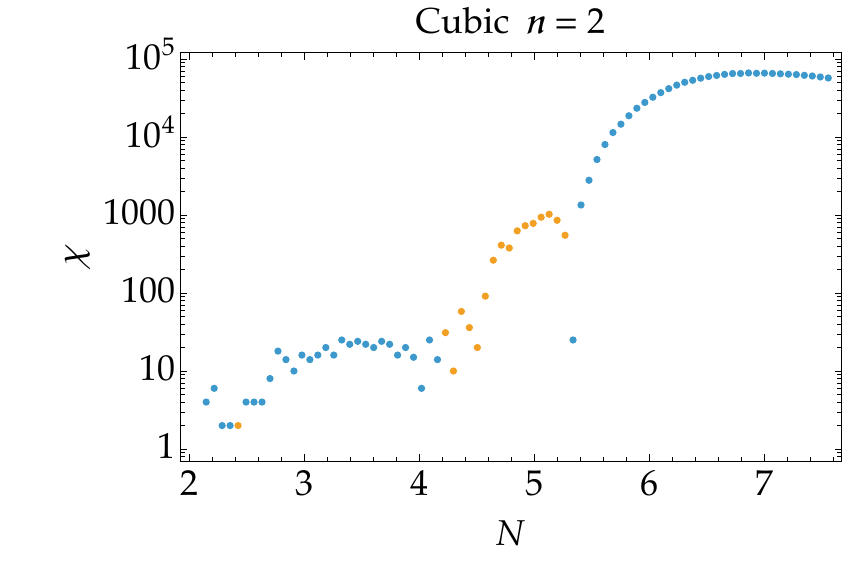}
        \end{minipage}
        \begin{minipage}{0.32\hsize}
            \centering
            \includegraphics[width=0.95\hsize]{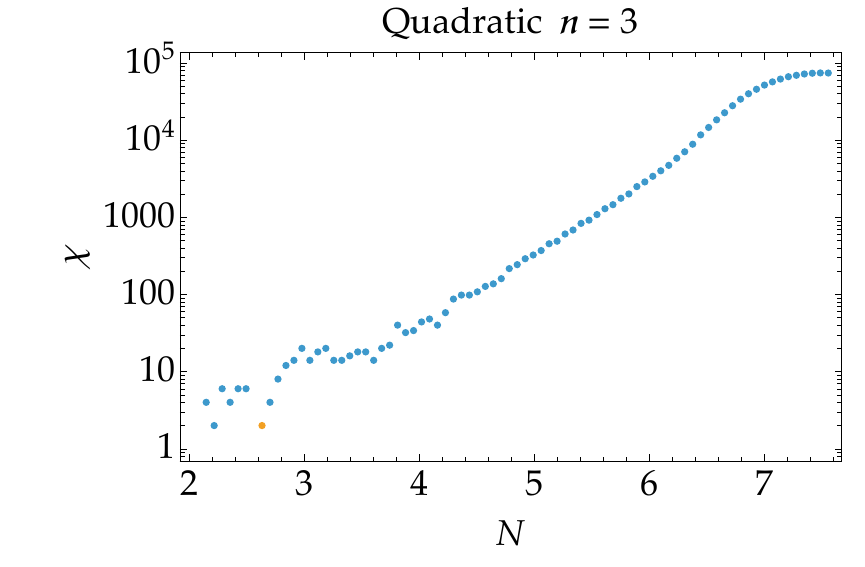}
            \includegraphics[width=0.95\hsize]{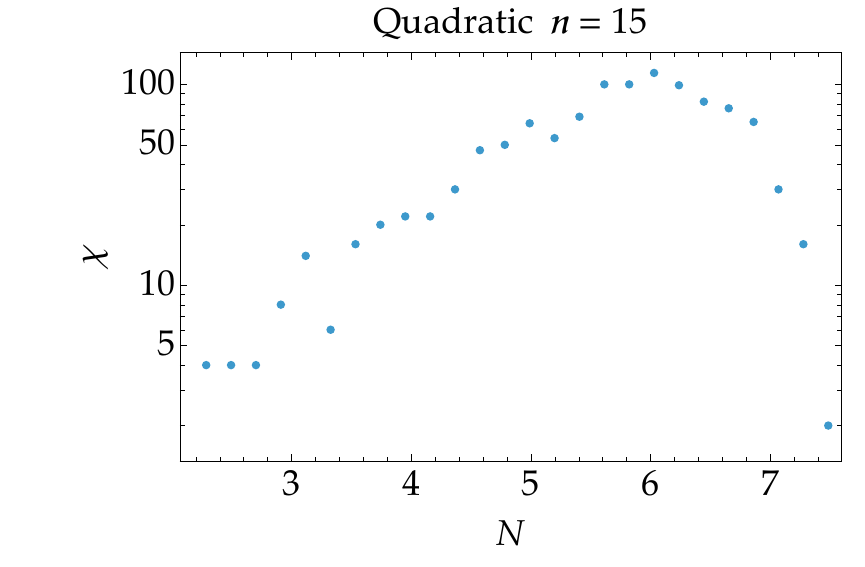}
        \end{minipage}
    \end{tabular}
    \caption{
    The total Euler characteristic $\chi$~\eqref{eq: chi total} of the waterfall fields below the threshold~\eqref{eq: psi threshold} as a function of the time $N$.
    The blue and orange dots show the positive and negative values, respectively.
    }
    \label{fig: EC}
\end{figure*}

In Fig.~\ref{fig: EC}, we plot the time-evolution of the Euler characteristic of the waterfall field for each model.
For $n=1$ or $2$, it becomes negative around $N\sim 4\text{--}5$, implying the emergence of holed domain walls or looped cosmic strings. It then turns into a large positive value because the stochastic noise reconnects the defects into finer structures and makes them unresolved, seen as isolated objects.
Its non-vanishment ensures the topologically stable structures.
Again, one finds that the topological structure does not significantly depend on the inflaton potential, i.e., ``Quadratic" or ``Cubic".

For $n=3$, the monopoles and their reconnections monotonically increase the Euler characteristic.
One can again see that the monopoles are stable as the Euler characteristic approaches a constant.
In the case of $n=15$, it first increases but then disappears because the structures are not topologically stable.

\subsection{Curvature perturbation}

\begin{figure*}
    \centering
    \includegraphics[width=0.7\hsize]{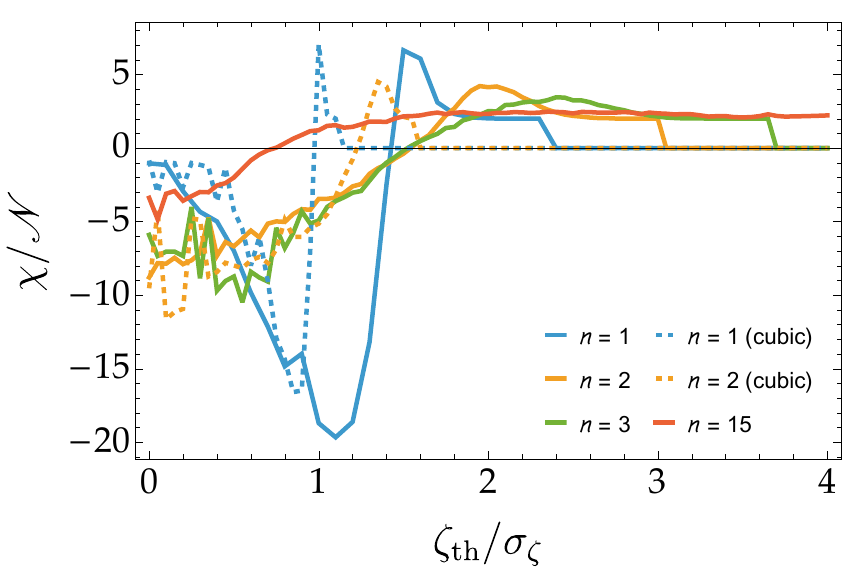}
    \caption{
    The Euler characteristic of the curvature perturbation divided by the number of objects in the simulation box, $\chi/\mathscr{N}$, evaluated for each threshold value $\zeta_\uth$ shown in the horizontal axis in units of the standard deviation $\sigma_{\zeta} =\sqrt{\overline{\delta\calN^2}}$.
    }
    \label{fig: EC_curvature}
\end{figure*}

The Euler characteristic can be calculated for the configuration of the final curvature perturbation as well.
In Fig.~\ref{fig: EC_curvature}, we show the Euler characteristic of the region where the curvature perturbation is smaller than a certain threshold value $\zeta_\uth$ as a function of $\zeta_\uth$. Its average over the number $\mathscr{N}$ of the connected objects is exhibited. It is calculated in a subregion of the size $64^3$ around the centre of the simulation box for numerical conciseness.
Only the case of $n=1$ shows significantly negative Euler characteristics for $\zeta_\uth\sim\sigma_\zeta$, where $\sigma_{\zeta}=\sqrt{\overline{\delta\calN^2}}$ is the standard deviation in the simulation box, implying global structures of the curvature perturbation.
We do not conclude whether it is because of the domain wall or the characteristic \ac{PDF} (it has a bump before the decay for large $\delta\calN$ as can be seen in Fig.~\ref{fig: pdf}), as the topological defects are basically unresolved in our simulations.
Indeed, the $n=2$ (cosmic string) case is not much different from the $n=3$ (monopole) case, contrary to the waterfall Euler characteristic shown in Fig.~\ref{fig: EC}.
For sufficiently high thresholds, the Euler characteristic approaches two, indicating that high peaks are configured isolatedly.
They finally disappeared for a too high threshold, and we define $\chi/\scrN=0$ if there is no corresponding object in the box.
In the ``Cubic" cases, as there are upper bounds on the curvature perturbation, the high-$\zeta$ region disappears even before it reaches the asymptotic value $\chi/\scrN\to2$, though the overall behaviours are similar to the counterpart in ``Quadratic".

\section{Conclusions}%
\label{sec: conclusion}

In this paper, we investigated the spatial profile of the curvature perturbation generated by a mild-waterfall-type hybrid inflation. 
Given that this model is known to produce topological defects, we performed simulations using \acf{STOLAS}, a lattice simulation code based on the stochastic formalism of inflation, to determine the topological configuration of the curvature perturbation that would be imprinted by defects.
We examined six cases (``Quadratic'' $n=1$, $2$, $3$, and $15$, and ``Cubic'' $n=1$ and $2$) by varying the number of waterfall fields $n$ and the functional form of the inflaton potential.

In Sec.~\ref{sec: PDF and PS}, we analysed the statistical properties of the curvature perturbation. 
Our results for the \acf{PDF} at $\delta\calN \gtrsim 1$ appeared to contradict previous research~\cite{Murata:2025onc}. 
This discrepancy arises because our results are averaged at each lattice point and thus coarse-grained at the lattice scale, while our previous computation corresponds to the minimal scale, i.e., the Hubble scale at the end of inflation. 
We expect that increasing the resolution will allow \ac{STOLAS} to reproduce the findings of~\cite{Murata:2025onc}, which leaves for future work.
For the ``Cubic'' case, Ref.~\cite{Murata:2025onc} previously pointed out the existence of the upper bound in the \ac{PDF}. 
We observed the same upper bound also in \ac{STOLAS}, and it results in a bunch of regions with $\zeta$ around the corresponding maximum value. 
In such configurations, regions exceeding $\delta\calN > 1$ do not exist; therefore, \acf{PBH} are seldom formed. 
These structures may affect halo formation, making future $N$-body simulations an intriguing direction for research.
It would also be interesting to investigate the scalar-induced gravitational wave (see, e.g., Refs.~\cite{Ananda:2006af,Baumann:2007zm,Saito:2008jc,Espinosa:2018eve,Kohri:2018awv,Domenech:2021ztg}) from such configurations of the curvature perturbation.

The power spectrum was derived through a direct Fourier transform of the curvature perturbation. 
Our results are well fitted using the analytical formula derived in Ref.~\cite{Tada:2023pue} based on the stochastic-$\delta\calN$ algorithm.
Except for deviations caused by slow-roll effects, the algorithm is consistent with the \ac{STOLAS} results.
Specifically, the discrepancies arise from the slow-roll approximation adopted in Ref.~\cite{Tada:2023pue}, where the field velocities $\pi_i$ are neglected. 
Indeed, Ref.~\cite{Miyamoto:2025qqm} recently computed the power spectrum for the ``Quadratic'' $n=1$ model within the narrow-sense stochastic-$\delta\calN$ approach keeping $\pi_i$, and found results quantitatively consistent with ours, confirming that the deviations are attributable to this approximation.
While the peak of the power spectrum is roughly proportional to $1/n$ in the ``Quadratic" case as known in the literature~\cite{Halpern:2014mca,Tada:2023fvd,Murata:2025onc}, we observed that the ``Cubic" case does not follow this relation because the amplitude of the power spectrum is mainly determined by the upper bound which does not depend on the waterfall number $n$.

In Sec.~\ref{sec: defect}, we employed the Euler characteristic to investigate the topological properties of the waterfall fields and curvature perturbation.
We found it consistent with the interpretation that topological defects emerge according to their respective symmetries at around the critical point, and then reconnect into finer and finer structures due to the stochastic noise, as can be seen directly in snapshots (Fig.~\ref{fig: snap}) as well.
We leave a detailed study about the correlation length of the defects in the mild-waterfall models for future work.
Despite the formation of the topological defects, it is suggested that only the case of $n=1$ leaves a non-trivial topological structure in the curvature perturbation.
The $n=1$ model suffers from the domain wall problem, but it could be solved if the two potential minima are not completely degenerate (see, e.g., Ref.~\cite{Saikawa:2017hiv} for a recent review).
Then, such a topological structure of the curvature perturbation may leave a unique signature in, e.g., the cosmological large-scale structure.

Since the curvature perturbation map generated by \ac{STOLAS} includes all information about the spatial correlation in principle, it can also calculate not only the power spectrum but also the bispectrum or higher-order correlators, for which no rigorous stochastic-$\delta\calN$ algorithm has been proposed.
It would also be important to investigate the practical \ac{PBH} scenario for $n\sim15$ with an appropriate \ac{PDF}.
We leave them for future work.

\acknowledgments
We are grateful to Takashi Hiramatsu for their helpful discussions.
This work is supported in part by the workstation facilities of the Institute of Theoretical Physics, Rikkyo University.
T.M. is supported by Specific Project Grant from TMCIT.
Y.T. is supported by JSPS KAKENHI Grant
No.~JP24K07047.

\appendix

\section{Computation of Euler characteristic in Mathematica}%
\label{app: EC in mathematica}

\begin{figure*}
    \centering
    \includegraphics[width=0.95\hsize]{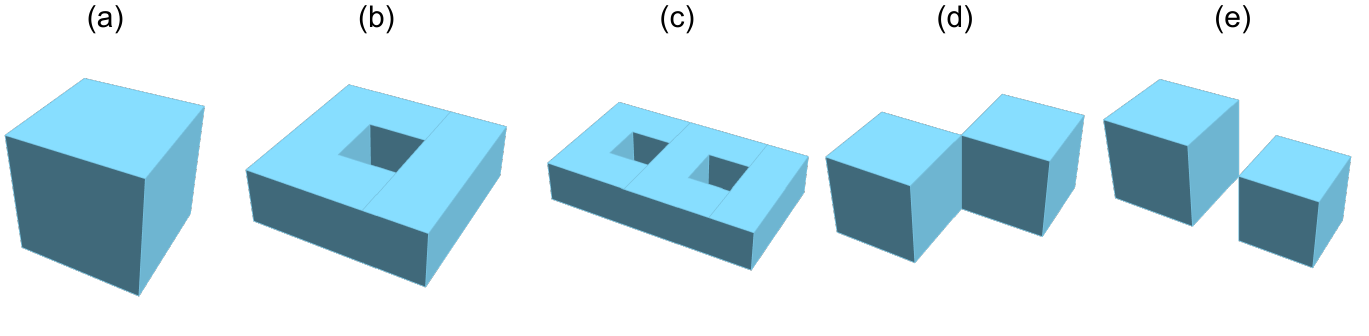}
    \caption{Some examples of the lattice-based structures.}
    \label{fig: EC_app}
\end{figure*}

We here summarise the specification in calculation of the Euler characteristic with Mathematica.
Each point in the lattice is given as a cube, and the Euler characteristic is calculated by counting the number of vertices $V$, edges $E$, and faces $F$ with $\chi = V-E+F$.
A single cube (case (a) in Fig.~\ref{fig: EC_app}) returns an Euler characteristic of $\chi = 8-12-6=2$.
An object containing one hole yields $\chi = 0$, while an object with two holes yields $\chi = -2$.
This corresponds to the general relation $\chi = 2 - 2g$, where $g$ is the genus (number of independent holes).
When two cubes share one full edge, Mathematica returns $\chi = 3$, due to the resulting counts: $V = 14,\ E = 23,\ F = 12$, which satisfy $\chi = V - E + F = 3$.
Even when the cubes share only one vertex, the Euler characteristic computed is again $\chi = 3$.
The value of geometric quantities for all cases in Fig.~\ref{fig: EC_app} is summarised in Table~\ref{tab: EC_app}.

\begin{table}
    \renewcommand{\arraystretch}{1.3}
    \centering
    \caption{Each case of geometric quantities and the Euler characteristic in Fig.~\ref{fig: EC_app}.}
    \label{tab: EC_app}
        \begin{tabular}{ccccc}
            \toprule
            case & $V$ & $E$ & $F$ & $\chi =V - E + F $
            \\
            \hline
            (a) & 8  & 12 & 6 & 2
            \\
            (b) & 20 & 32 & 12 & 0
            \\
            (c) & 32 & 52 & 18 & $-2$
            \\
            (d) & 14 & 23 & 12 & 3
            \\
            (e) & 15 & 24 & 12 & 3 \\
            \bottomrule
        \end{tabular}
\end{table}

\section{The $\sigma$ dependence of the results}\label{app:sigma}

\begin{figure}
    \centering
    \begin{tabular}{c}
        \includegraphics[width=0.7\hsize]{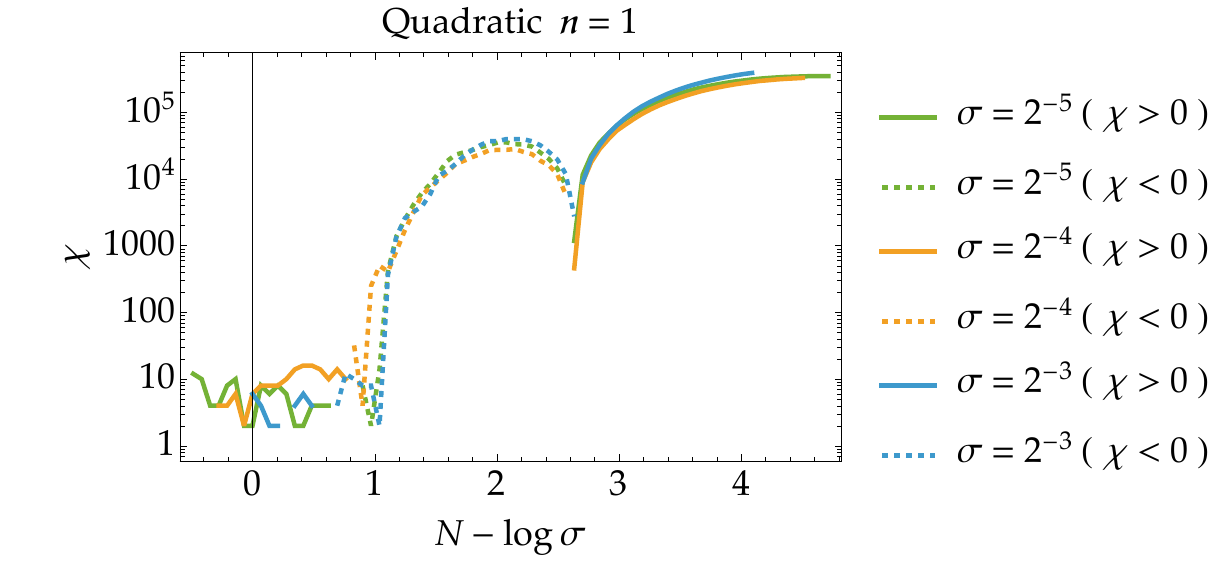}
        \end{tabular}
    \caption{
    The $\sigma$ dependence of the Euler characteristic $\chi$ as a function of $\log n_\sigma=N-\log\sigma$ in the ``Quadratic" $n=1$ model, similarly to Fig.~\ref{fig: EC}.
    The solid and dashed lines show the positive and negative values, respectively.
    The blue, orange, and green lines correspond to the value of $\sigma=2^{-3}$, $2^{-4}$, and $2^{-5}$, respectively.
    }
    \label{fig:sigma_euler}
\end{figure}

To check the dependence of our results on the coarse-graining parameter $\sigma$, we perform simulations for the ``Quadratic'' $n=1$ model with $\sigma = 2^{-3}$, $2^{-4}$ (the main text choice), and $2^{-5}$, with the same grid size $N_L = 256$ as in the main text.
The results are shown in Fig.~\ref{fig:sigma_euler}, where we compare the Euler characteristic of the radial direction of the waterfall fields for the three values of $\sigma$.
One finds that the results are insensitive to the choice of $\sigma$ within this range, validating the gradient expansion and the stochastic formalism we follow.

\section{Zoom-in method in STOLAS}\label{sec: zoom in}

In our STOLAS, the spatial zoom-in method offers an advantage compared to conventional lattice simulations (see, e.g., Refs.~\cite{Figueroa:2020rrl,Caravano:2021pgc,Caravano:2025klk}). 
In the stochastic formalism, subHubble quantum fluctuations are coarse-grained, and appear as classical stochastic noise to the superHubble modes at each time step.
Furthermore, the spatial derivative is neglected in the \ac{EoM}~\eqref{eq: discrete EoM} as a leading-order gradient expansion; the spatial correlation appears only through the noise correlation~\eqref{eq: DW correlation}.
Therefore, zooming in on a specific spatial sub-volume by inserting new lattice points among grids at a certain time step does not cause any harmful backreaction onto the original grids.
The zoom-in method allows us to
track the simulation down to smaller scales.
In conventional lattice simulations, where all relevant fluctuations are injected at the beginning of the simulation and the lattice points are nontrivially coupled via the spatial derivatives, it is basically impossible to compute, e.g., the power spectrum at frequencies higher than the original Nyquist frequency.

For the zoom-in simulations, we refine the lattice grid by a factor of two in each spatial dimension.
The zoom-in procedure is triggered when the dimensionless wavenumber satisfies $n_{\sigma}=16$.
Specifically, we select the central region of the simulation box and extract a localised sub-volume of size $(N_L/2)^3$.
Next, we reconstruct a fine grid with the same number of points $N_L^3$ within this localised volume. 
The field values on the new fine grid are determined via trilinear interpolation from the coarse sub-grid fields.

\begin{figure*}
    \centering
    \begin{tabular}{c}
        \includegraphics[width=0.7\hsize]{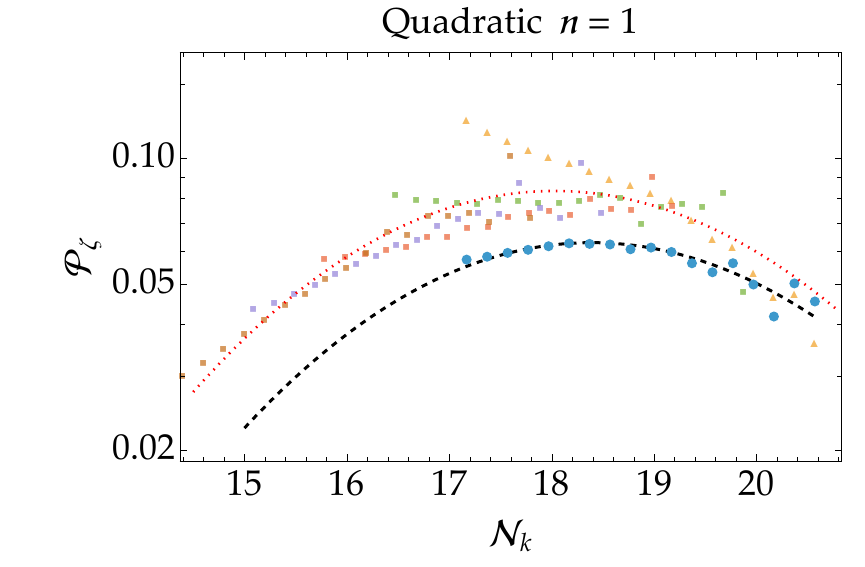}
        \end{tabular}
    \caption{
    Plot of the power spectra in the zoom-in method and the averaging procedure.
    Blue dots and the black dashed line indicate the averaging results mentioned in the main text.
    The colour triangles are the results without any procedure (i.e., before the zoom-in), and squares represent the results for each slice when the zoom-in method is used.
    The red dotted line shows the analytic formula~\eqref{eq:analytic_formula} with fitting parameters $(N_{\rm water}, \calP_{\zeta}^{\rm peak})= (18.1,\ 8.34\times 10^{-2})$.}
    \label{fig:power_zoom}
\end{figure*}

We apply this method to the ``Quadratic" $n=1$ model with a four-times zoom-in. 
In Fig.~\ref{fig:power_zoom}, we show the power spectra of the curvature perturbation for individual slices, without the averaging procedure, alongside the averaged result presented in the main text. 
Although the averaging procedure results in a slightly smaller power spectrum, the overall shape and qualitative behaviour are consistent between the two, confirming that the averaging procedure captures the correct qualitative features.

\bibliographystyle{JHEP}
\bibliography{main_STOLAS}
\end{document}